\documentclass[11pt,a4paper]{article}
\usepackage[utf8]{inputenc}
\usepackage{amssymb}
\usepackage{amsmath}
\usepackage{url}
\usepackage{graphicx}
\usepackage{subcaption}
\usepackage{natbib}
\usepackage[left=3cm,right=3cm,top=2.5cm,bottom=2.5cm]{geometry}
\usepackage{comment}
\usepackage[]{algorithm}
\usepackage{algpseudocode}
\usepackage{color}

\title{Stratified sampling and bootstrapping for approximate Bayesian computation}
\author{Umberto Picchini\footnote{Department of Mathematical Sciences, Chalmers University of Technology and the University of Gothenburg, Sweden, picchini@chalmers.se.} \and Richard G. Everitt\footnote{Department of Statistics, University of Warwick, UK}
}

\date{}

\begin{document}
\maketitle

\begin{abstract}
Approximate Bayesian computation (ABC) is computationally intensive for complex model simulators. To exploit expensive simulations, data-resampling via bootstrapping can be employed to obtain many artificial datasets at little cost. However, when using this approach within ABC, the posterior variance is inflated, thus resulting in biased posterior inference. Here we use stratified Monte Carlo to considerably reduce the bias induced by data resampling. We also show empirically that it is possible to obtain reliable inference using a larger than usual ABC threshold. Finally, we show that with stratified Monte Carlo we obtain a less variable ABC likelihood. Ultimately we show how our approach improves the computational efficiency of the ABC samplers. We construct several ABC samplers employing our methodology, such as rejection and importance ABC samplers, and ABC-MCMC samplers. We consider simulation studies for static (Gaussian, g-and-k distribution, Ising model, astronomical model) and dynamic models (Lotka-Volterra). We compare against state-of-art sequential Monte Carlo ABC samplers, synthetic likelihoods, and likelihood-free Bayesian optimization.  For a computationally expensive  Lotka-Volterra case study, we found that our strategy leads to a more than 10-fold computational saving, compared to a sampler that does not use our novel approach.  
\end{abstract}

\section{Introduction}\label{sec:introduction}

The use of realistic models for complex experiments typically results in an intractable likelihood function, i.e. the likelihood is not analytically available in closed form, or it is computationally too expensive to evaluate. 
Approximate Bayesian computation (ABC) is arguably the most popular simulation-based inference method \citep{cranmer2019}, and is sometimes denoted ``likelihood-free'' inference (recent reviews are \citealp{lintusaari2017fundamentals} and \citealp{karabatsos2017approximate}). The key feature of ABC is the use of models as ``computer simulators'', these producing artificial data that can be used to compensate for the unavailability of an explicit likelihood function. The price to pay for the flexibility of the approach, only requiring forward simulation from the simulators, is that the resulting inference is approximate and the computational requirements are usually non-negligible. 

In this work we consider ways to alleviate the computational cost of running ABC with computationally expensive simulators, by using resampling of the artificial data (i.e. non-parametric bootstrap) to create many artificial samples that are used to approximate the ABC likelihood function. This idea has previously been used in \cite{bootstrapped-sl}, primarily in the context of inference via ``synthetic likelihoods'' \citep{wood2010statistical,price2018bayesian}. However we have noticed that when this approach is used within ABC it produced biased inference, with resulting marginal posteriors having very heavy tails. 

The goal of our work is to construct a procedure that alleviates the bias produced by data-resampling. We achieve our goal using stratified Monte Carlo. Interestingly, we also show that, thanks to  stratified Monte Carlo embedded in a bootstrapping approach, we can produce reliable ABC inference while keeping the so-called ABC threshold/tolerance to a relatively large value. We show how to use a larger threshold to obtain a better mixing chain with a relatively high acceptance rate, and hence reduce the number of iterations, without compromising on the quality of the inference. We apply our strategies to ABC-MCMC, rejection-ABC and importance sampling ABC. In an example where stratified Monte Carlo is applied to both a rejection ABC algorithm and to importance sampling ABC (see section \ref{sec:supernova}), we show that we require two to four times fewer model simulations (depending on the method and the setup) when using stratified Monte Carlo. We also obtain an 11 to 13-folds acceleration on an expensive Lotka-Volterra model when applying stratified Monte Carlo to ABC-MCMC. Note that in \cite{andrieu2016establishing} there is a theoretical discussion on the general benefits of using stratification into ABC. They briefly suggest possibilities to accelerate ABC sampling using some kind of ``early rejection'' scheme, however this only applies to very simple examples, and specifically to cases where summary statistics are not used, which is not a typical scenario.
We instead give a practical and more general construction of ABC algorithms that exploit stratification, with emphasis on the use of resampling methods. 

The paper is organized as follows. In section \ref{sec:intro-abc} we introduce basic notions of ABC. In section \ref{sec:sampling-resampling-stratification} we discuss stratified Monte Carlo. In section \ref{sec:abc} we show how to approximate an ABC likelihood using stratified Monte Carlo. In section \ref{sec:abc-mcmc} we construct an ABC-MCMC sampler using bootstrapping (resampling) and stratified Monte Carlo.
Section \ref{sec:case-studies} considers four case studies. A Supporting Material section illustrate details about our case studies, consider a further case study (g-and-k model) and details an importance sampling ABC algorithm using stratified Monte Carlo as well as the rationale to identify strata for the case studies. Code can be found at \url{https://github.com/umbertopicchini/stratABC}.

\section{Approximate Bayesian computation}\label{sec:intro-abc}

We are interested in inference for model parameters $\theta$, given observed data $x$, and we denote the likelihood function for the stochastic model under study with $p(\cdot|\theta)$, then $x\sim p(\cdot|\theta^o)$ for some parameter value $\theta=\theta^o$. The key aspect of any simulation-based inference methodology is to avoid the evaluation of a computationally intractable $p(\cdot|\theta)$, and instead rely on simulations of artificial data generated from a model, the (deterministic or stochastic) \textit{simulator}. A simulator is essentially a computer program, which takes $\theta$, makes internal calls to a random number generator (if the simulator is stochastic), and outputs a vector of artificial data.
If we denote with $x^*$ artificial data produced by a run of the simulator, conditionally on some parameter $\theta^*$, then we have that $x^*\sim p(\cdot|\theta^*)$.  In a Bayesian setting, the goal is to sample from the posterior distribution $\pi(\theta|x)$, however this operation can be impossible or at best challenging, since the posterior is proportional to the possibly unavailable likelihood via $\pi(\theta|x)\propto p(x|\theta)\pi(\theta)$, with $\pi(\theta)$ the prior of $\theta$. 
ABC is most typically constructed to target a posterior distribution $\pi_\delta(\theta|s)$, depending on summary statistics of the data $s$, where $s=S(x)$ for $S(\cdot)$ a function $S:\mathbb{R}^{n_\mathrm{obs}}\rightarrow \mathbb{R}^{n_s}$ with $n_\mathrm{obs}$ the number of data points, and $n_s=\dim(s)$, and finally a tolerance (threshold) $\delta>0$. We discuss the essential properties of ABC in section \ref{sec:abc} but here we anticipate that if $s$ is informative for $\theta$, then $\pi_\delta(\theta|s)\approx \pi(\theta|x)$ for small $\delta$, thus allowing approximate inference. Denote with $s^*$ the vector of summary statistics associated to the output $x^*$ produced by a run of the simulator, i.e. $s^*=S(x^*)$. Then we have an approximate posterior 
\[\pi_\delta(\theta|s)\propto \pi(\theta)\int \mathbb{I}_{\{\parallel s^*-s\parallel<\delta\}} p(s^*|\theta)ds^*\]
for some distance $\parallel\cdot \parallel$ and indicator function $\mathbb{I}$. The smaller the value of $\delta$, the more accurate the approximation to $\pi(\theta|s)$. More generally, we will consider
$\pi_\delta(\theta|s)\propto \pi(\theta)\int K_\delta(s^*,s) p(s^*|\theta)ds^*$
for an unnormalised density (kernel function) $K_\delta(s^*,s)$. 
ABC algorithms can be implemented in practice through the pseudo-marginal approach using, for each $\theta$, a Monte Carlo estimate of the integral (commonly known as the ``ABC likelihood'') in the previous expression:
\begin{equation}
\int K_\delta(s^*,s) p(s^*|\theta)ds^*\approx \frac{1}{M}\sum_{r=1}^M K_\delta(s^{*r},s), \qquad s^{*r}\sim_{iid} p(s^*|\theta), \quad r=1,...,M.\label{eq:abc-likelihood}
\end{equation}
The choice of $M$ goes some way to dictating the efficiency of the method: a larger $M$ gives lower variance estimates of the likelihood and would generally improve the statistical efficiency of ABC algorithms. However, the choice $M=1$ is often favoured: in most cases because this was the approach originally described in the ABC literature (e.g. \citealp{marjoram2003markov}); but also because the increased computational cost of taking $M>1$ often outweighs the improvement in efficiency \citep{bornn2017use}. As an example of an algorithm taking $M>1$, the generalized version of the ABC-rejection algorithm \citep{bornn2017use} is given in algorithm \ref{abc-gen}.

\begin{algorithm}
\scriptsize
\begin{algorithmic}[1]
  \For{$t=1$ to $N$}
    \Repeat 
   \State Generate $\theta^{*} \sim \pi(\theta)$,
     $x^{i*} \sim p(x| \theta^{*})$ i.i.d. for $i=1,\dots, M$;
   \State compute $s^{i*}=S(x^{i*})$, $i=1,\dots, M$;
   \State simulate $u \sim \mbox{Uniform}(0,1)$;
    \Until{$u <  \frac{1}{cM} \sum_{i=1}^M K_\delta(s^{i*},s)$}
   \State Set $\theta^{(t)} = \theta^*$\;
  \EndFor
  \end{algorithmic}
\caption{Generalized ABC-rejection. Here $c$ is a constant satisfying $c \geq \sup_z
K_\delta(z)$. When using a Gaussian kernel as in this paper, we can take $c=1$. \label{abc-gen}}
\end{algorithm}

We construct versions of ABC algorithms that, when the simulator is computationally intensive, are fast compared to using $M\gg 1$, by employing resampling ideas, to approximately target the true summaries-based posterior $\pi(\theta|s)$. We will only ``approximately'' target the latter, as in addition to having $\delta>0$, the resampling procedure introduces an additional source of variability that biases the posterior, that is it produces an ABC posterior having a larger variance. We consider stratified Monte Carlo to reduce the bias due to resampling, and show that coupling resampling with stratified Monte Carlo produces inference that more closely approximates the true $\pi(\theta|s)$ (compared to using resampling without stratification). We show that in this case we are able to use a larger ABC threshold $\delta$ than typically required, while still obtaining accurate inference. We now introduce stratified Monte Carlo.

\section{Bootstrapping and stratified Monte Carlo}\label{sec:sampling-resampling-stratification}

Consider the problem of approximating the following integral 
\begin{equation}
\mu = \int_{\mathcal{D}}f(z)p(z)dz \label{eq:mu}
\end{equation}
over some space $\mathcal{D}$,
for some function $f$, density function $p$, and a generic variable $z$. 
Later $\mu$ will represent a likelihood function, but for the moment we cast the problem into a general framework. Clearly we can approximate $\mu$ using Monte Carlo, i.e. by generating multiple independent samples $z^r$ from $p(z)$ ($r=1,...,M$) we compute
\begin{equation}
\hat{\mu}=\frac{1}{M}\sum_{r=1}^M f(z^r), \qquad z^r\sim_{iid} p(z), \quad r=1,...,M.\label{eq:standard-mc}
\end{equation}
However this can be computationally expensive, if $p(\cdot)$ represents the probabilistic structure of a complex stochastic simulator, and we could instead simulate only few times from $p(z)$. As an illustration, suppose $z$ is a vector of independent and identically distributed (iid) elements of length $n_\mathrm{obs}$. Our approach is then to simulate a single vector $z\sim p(z)$ and resample $n_\mathrm{obs}$ times with replacement from the $n_\mathrm{obs}$ elements of $z$, to obtain a pseudo-sample $z^{*1}$, hence $\dim(z^{*1})=\dim(z)$. We repeat the resampling procedure on the same $z$ further $R-1$ times, so in the end we have the vectors $z^{*1},..,z^{*R}$, each having $\dim(z^{*r})=n_\mathrm{obs}$ (in practice this is a non-parametric bootstrap approach returning $R$ samples). Then we define the ``bootstrapped'' estimator $\hat{\mu}_{\mathrm{res}}$ obtained using resampling as
\begin{equation}
\hat{\mu}_{\mathrm{res}}=\frac{1}{R}\sum_{r=1}^R f(z^{*r}).\label{eq:resampled-mc}
\end{equation}
For case studies where data are assumed iid, we always use uniform resampling, though alternatives can be considered (dependent data and multivariate time series with several resampling approaches are also considered in our work).
While the resampling approach can be much faster than using \eqref{eq:standard-mc} to obtain an estimate of $\mu$ (or at least this is true when simulating from $p(x)$ is computationally expensive), the problem is that when $\hat{\mu}_{\mathrm{res}}$ is used as an estimator of the ABC likelihood, the resulting posterior distribution is overdispersed. This finding was already discussed in \cite{bootstrapped-sl}. Also, using \eqref{eq:resampled-mc} to approximate the ABC likelihood results in a very biased estimate of $\mu$, producing a posterior with a large variance. This is shown in section \ref{sec:biased-likelihood}.

When data are not iid, there is a large literature for resampling schemes for dependent data. We do not go into details and refer the interested reader to review papers such as \cite{hardle2003bootstrap}, \cite{kreiss2011bootstrap} or the  monography \cite{lahiri2013resampling}. In section \ref{sec:lv} we consider a time-series which we resample using different bootstrap schemes: the standard block-bootstrap, with and without overlapping blocks, and the stationary bootstrap.
For example, the block bootstrap \citep{kunsch1989jackknife}
resamples blocks of data. These blocks are chosen to be sufficiently
large such that they retain the short range dependence structure of
the data, so that a resampled time series constructed by concatenating
resampled blocks has similar statistical properties to a real sample. Suppose that $x_{1:n_{_\mathrm{obs}}}\sim p(x)$ is a vector of time-indexed data. In the block bootstrap,
using a block of length $B$ (for simplicity we consider the case
where $B$ is a divisor of $n_\mathrm{obs}$), we may construct a set of overlapping or non-overlapping
blocks of indices of the observations. An example of overlapping blocks is
\begin{equation*}
\mathcal{B}=\left\{ (1:B),(2:B+1),...,(n_\mathrm{obs}-B+1:n_\mathrm{obs})\right\} .
\end{equation*}
Then a resample $x^r$ from $x_{1:n_\mathrm{obs}}$ consists
of $n_\mathrm{obs}/B$ concatenated blocks whose indices are sampled with replacement
from the ``blocks of indeces'' $\mathcal{B}$. The summary statistics of $R$ resamples $\left\{ x^r\right\} _{r=1}^{R}$
may then be computed. Experiments with several bootstrap methods for observations from a Lotka-Volterra model have also been detailed in Supplementary Material.

\subsection{Stratified Monte Carlo}\label{sec:stratified-sampling}

We wish to obtain variance reduction via stratified Monte Carlo (e.g. \citealp{rubinstein2016simulation}). We partition the integration space $\mathcal{D}$ (see \eqref{eq:mu}) into  $J$ ``strata'' $\mathcal{D}_j$, and the resulting estimator of $\mu$ is 
\begin{equation}
\hat{\mu}_{\mathrm{strat}}= \sum_{j=1}^J\frac{\omega_j}{\tilde{n}_j}\sum_{i=1}^{\tilde{n}_j}f(x_{ij}).\label{eq_mu-strat}
\end{equation}
Here $\omega_j$ are known probabilities, with $\omega_j=\mathbb{P}(X\in \mathcal{D}_j)$, $\tilde{n}_j$ is the number of Monte Carlo draws that the experimenter decides to sample from stratum $\mathcal{D}_j$, and $x_{ij}$ is the $i$th draw generated from the model and within stratum $\mathcal{D}_j$, with $\dim(x_i)=\dim(x)$. That is, in this case knowledge is assumed of how to directly generate draws from each stratum. Under the stringent condition that the $\omega_j$ are known, it can be easily shown that $\hat{\mu}_\mathrm{strat}$ is an unbiased estimator of ${\mu}$ (e.g. \citealp{owen}, chapter 8; this is also given as Supplementary Material for ease of access). However, in the following we are not assuming that the $\omega_j$ are known (nor that we are able to simulate from a given stratum), and show how to proceed to their estimation. We assume an approach similar to the ``post stratification'' in \cite{owen}, meaning that we sample $x_{i}\sim p(x)$ with $x\in \mathcal{D}$,
and assign each $x_i$ to one of the strata ``after the fact'', as opposed to sampling directly from a given stratum (the latter would be ideal but also not a readily available approach). The difference with the actual post-stratification is that in our case the $\omega_j$ have to be estimated, whereas in the original post-stratification the $\omega_j$ are known. 
A consequence of using post-stratification is that, by defining with $n_j$ the cardinality of the set $\{i;x_i\in \mathcal{D}_j\}$, then $n_j$ is a random variable (hence the notational difference from the $\tilde{n}_j$ in \eqref{eq_mu-strat}). Therefore the value of $n_j$ is known \textit{after} the simulation is performed, while $\tilde{n}_j$ is set beforehand by the experimenter. In practice we will use the following estimator
\begin{equation}
\hat{\hat{\mu}}_{\mathrm{strat}}= \biggl\{\sum_{j=1}^J\frac{\hat{\omega}_j}{{n}_j}\sum_{i=1}^{{n}_j}f(x_{ij})\biggr\}\mathbb{I}_{\{n_j>0,\forall j\}}=
\begin{cases}
\sum_{j=1}^J\frac{\hat{\omega}_j}{{n}_j}\sum_{i=1}^{{n}_j}f(x_{ij}),\qquad \text{if } n_j>0,\forall j\\
0,\qquad \text{otherwise,}
\end{cases}    
,\label{eq_mu-strat-biased}
\end{equation}
where $\mathbb{I}_{\{n_j>0,\forall j\}}$ is the indicator function equal to one when all $n_j$ are positive and zero otherwise.
Notice the double ``hat'' since this estimator uses estimated strata probabilities $\hat{\omega}_j$, whose construction is detailed in section \ref{sec:train-test}. In \eqref{eq_mu-strat-biased} we impose the estimator to be zero as soon as $n_j=0$ for some $j$. This is only necessary when \eqref{eq_mu-strat-biased} is used in an inference algorithm, and in this case a parameter proposal is immediately rejected as soon as a stratum is neglected (i.e. $n_j=0$ for some $j$). We will show how this property has both downsides and upsides. The upside is that we only evaluate proposals for which the integral of the corresponding ABC likelihood is approximated using samples that cover \textit{all strata}, including the most internal stratum. This has the major benefit of producing a less variable ABC likelihood approximation, since the approximation to the integral in \eqref{eq:mu} is strictly positive only when all strata have been ``hit''.  A downside is of course that a single neglected stratum causes immediate rejection. However, we alleviate the latter issue by producing many samples at a small cost using bootstrapping as described in section \ref{sec:sampling-resampling-stratification}. 

A consequence of estimating the probabilities $\omega_j$ is that $\hat{\hat{\mu}}_{\mathrm{strat}}$ is not unbiased, unlike \eqref{eq_mu-strat}. In fact, if we denote with $x^*_j$ the sequence of draws $x^*_j=(x_{1j},...,x_{n_jj})$ ending in $\mathcal{D}_j$, then $n_j$ is depending on this sequence, i.e. $n_j\equiv n_j(x^*_j)$, and

\begin{equation}
\mathbb{E}(\hat{\hat{\mu}}_{\mathrm{strat}})=
\begin{cases}
\sum_{j=1}^J \mathbb{E}(\frac{\hat{\omega}_j}{{n}_j(x_j^*)}\sum_{i=1}^{{n}_j}f(x_{ij})),\qquad \text{if } n_j(x_j^*)>0,\forall j\\ \label{eq:biased-expectation}
0,\qquad \text{otherwise.}
\end{cases}    
\end{equation}
Since our framework assumes that $p(x)$ itself is unknown, and that we only know how to sample from it, it turns out that the distribution of $n_j(x^*_j)$ is unknown, and that  $\mathbb{E}(\hat{\hat{\mu}}_{\mathrm{strat}})$ is intractable. We reconsider again this expression in equation \eqref{eq:biased-expectation-2}, after having introduced our ABC methodology.

\section{ABC using stratification}\label{sec:abc}

In ABC we consider the posterior $\pi_\delta(\theta|s)\propto  \pi(\theta)\mu_\delta(\theta) = \pi(\theta)\int K_\delta(s^*,s) p(s^*|\theta)ds^*$, with $\mu_\delta(\theta)=\int K_\delta(s^*,s) p(s^*|\theta)ds^*$ representing the ``ABC likelihood''. An unbiased estimator of the ABC likelihood is given by
\begin{equation}
    \hat{\mu}_\delta(\theta)=\frac{1}{M}\sum_{r=1}^M K_\delta(s^{*r},s),\qquad s^{*r}\sim_{iid}p(s|\theta),\quad r=1,...,M. \label{eq:pmABC-lik}
\end{equation}
As alluded to previously, \cite{bornn2017use} show that choosing $M=1$ is usually close to optimal. Specifically, they prove that, for indicator kernels, using $M=1$ yields a running time within a factor of two of optimal. This means that, although likelihood estimators obtained with $M=1$ necessarily have higher variance, these come with a small enough computational cost that makes the tuning of $M$ not worth the additional computational cost of simulating multiple times from the model.
We consider whether it is instead worth to make use of a large number of ``bootstrapped datasets'', which is the $R$ in section \ref{sec:sampling-resampling-stratification} (say $R=500)$. In the suggested approach we only run the model simulator once ($M=1$) or twice ($M=2$) per iteration. 

To lighten the notation, here and in the following we write $\mu(\theta)$ instead of $\mu_\delta(\theta)$, that is $\mu(\theta)$ never represents the true likelihood, and instead it is the ABC likelihood $\mu_\delta(\theta)$.
With reference to the notation in \eqref{eq:mu}, here we have $f(\cdot)\equiv K_\delta$, and take $\mathcal{D}\equiv \mathbb{R}^{n_s}$. We now consider the use of stratified sampling in this context. 
We first illustrate stratification when using an indicator kernel, and show that this would not be an appropriate choice. 
Consider the ABC kernel $K_\delta(s^*,s)=\mathbb{I}_{\parallel s^*-s\parallel<\delta}$. Suppose we partition $\mathcal{D}$ using two strata $\mathcal{D}_1$ and $\mathcal{D}_2$, with $\mathcal{D}_1=\{s^* \text{ s.t. } \parallel s^*-s\parallel<\delta\}$ and $\mathcal{D}_2=\mathcal{D}\backslash\mathcal{D}_1$ where $K_\delta$ equals 1 for every $s^*\in \mathcal{D}_1$ and equals 0 for every $s^*\in \mathcal{D}_2$. Clearly the ABC likelihood $\mu(\theta)=\int \mathbb{I}_{\{\parallel s^*-s\parallel<\delta\}} p(s^*|\theta)ds^*$ is approximated via stratified sampling as
\[
\hat{\mu}_{\mathrm{strat}}= \frac{\omega_1}{\tilde{n}_1}\sum_{i=1}^{\tilde{n}_1}1 + \frac{\omega_2}{\tilde{n}_2}\sum_{i=1}^{\tilde{n}_2}0=\omega_1.
\]
And here comes the problem that the strata probabilities $\omega_j$ are generally unknown.
We proceed to the estimation of $\omega_j$ using a second, independent simulation round. This is the ``post-stratification'' mentioned in section \ref{sec:stratified-sampling}, implying the inability to sample conditionally on strata, i.e. sample from $p_j(s^*|\theta)=p(s^*|s^*\in\mathcal{D}_j,\theta)$ and instead sample from $p(s^*|\theta)$. Since $\omega_j=\int_{\mathcal{D}_j} p(s^*|\theta)ds^*$, this can be approximated using say $M$ simulations from $p(s^*|\theta)$ (or, as we do in practice, produce a single simulation from $p$, then resample this $R$ times) so that $\hat{\omega}_1=\# \{s^* \in \mathcal{D}_1\}/M$ which implies
$\hat{\hat{\mu}}_{\mathrm{strat}}=\hat{\omega}_1=\sum \mathbb{I}_{\parallel s^*-s\parallel<\delta} /M$. This is the ABC likelihood that we would intuitively obtain via Monte Carlo  (with biased variance if we use resampling). Therefore, by using stratification with an indicator kernel and unknown $\omega_j$, we have not learned anything new, as we just recovered the standard Monte Carlo estimator. And as we show in some of our case studies, when ABC is coupled to a resampling strategy, the resulting inference is largely suboptimal. We can generalize the example above to more strata and reach the same conclusion.

However, we can just use a different (non-flat) ABC kernel, for example the Gaussian kernel 
\[K_\delta(s^*,s)=\frac{1}{\delta^{n_s}}\exp(-\frac{1}{2\delta^2}(s^*-s)'\Sigma^{-1}(s^*-s))\]
for $s\in \mathbb{R}^{n_s}$ (there $\Sigma$ is a $n_s\times n_s$ matrix normalizing the contributions of the components of $s$, typically obtained from a pilot run of ABC). This is a choice that  worked well for our case studies. Using a kernel other than the indicator one, we can express the stratified estimator so that the $\hat{\omega}_j$ are ``weighted'' by the $n_j$ (see \eqref{eq_mu-strat-biased}). That is $n_j$ will not cancel-out with the factor $\sum_{i=1}^{n_j}K_\delta(s^*_{ij},s)$, which is instead the case when $K_\delta$ is a flat kernel. 

As an example, for a Gaussian kernel we could define $\mathcal{D}$ to be partitioned into three strata, say 
$\mathcal{D}_1=\{s^* \text{ s.t. } \sqrt{(s^{*}-s)'\Sigma^{-1}(s^{*}-s)}\in (0,\delta/2]\}$, $\mathcal{D}_2=\{s^* \text{ s.t. } \sqrt{(s^{*}-s)'\Sigma^{-1}(s^{*}-s)}\in (\delta/2,\delta]\}$, $\mathcal{D}_3=\{s^* \text{ s.t. } \sqrt{(s^{*}-s)'\Sigma^{-1}(s^{*}-s)}\in (\delta,\infty)\}$.
 Therefore $\mathcal{D}_1$ is the stratum where the integrand is most ``important'' ($K_\delta$ has higher density values though not necessarily most of the mass), $\mathcal{D}_2$ is less important than $\mathcal{D}_1$ but more important than $\mathcal{D}_3$. This implies that, when we use an ABC kernel having infinite support, e.g. a Gaussian kernel, and when $\delta$ is small, the last stratum $\mathcal{D}_J$ will be the one receiving the largest number of draws $n_J$, and likely it will be $n_J\gg n_j$ for all $j=1,...,J-1$. Notice, the above is just an exemplification and by no means the edges of the strata have to depend on $\delta$. The edges can be arbitrary, though we show some explicit construction in the Supplementary Material. Briefly, we can produce some preliminary simulated datasets from the prior predictive distribution, then we bootstrap the simulated data and compute (cheaply) many simulated distances of the type $d=\sqrt{(s^{*}-s)'\Sigma^{-1}(s^{*}-s)}$. From these many distances, we can then compute small quantiles from the empirical distribution of the simulated distances, and use these quantiles to define the edges of the strata. This procedure does not really add much in term of computational effort, compared to methods not using stratified Monte Carlo. In fact, realistically, in ABC the user often needs to produce several pilot runs, at the very least to assess the variation of each of the $n_s$ elements that compose the vector of summaries (see the matrix $\Sigma$), and that allows an appropriate normalization of both the observed summaries and the simulated ones. Summaries simulated in such preparatory runs can be recycled (after normalization) to compute quantiles of the distances, and hence obtain information to produce the strata would be a simple by-product of preparatory (and typically necessary) runs. Preparatory runs could also be necessary for other reasons, say to construct summary statistics ``semi-automatically'', as in \cite{fearnhead-prangle(2011)} and \cite{wiqvist2019partially}.
 
\subsection{Construction of strata probabilities for ABC}\label{sec:train-test}
 
Here we suggest a way to obtain the frequencies $n_j$ and to estimate the probabilities $\omega_j$.  Notice, these quantities are actually dependent on $\theta$, however, for simplicity of notation we drop the reference to $\theta$ in the following.

\paragraph{$n_j$ determination:}
For given $\theta$  we generate, using resampling with replacement, $R$ samples $s^{*1},...,s^{*R}$ as follows: we simulate a single $x^*$ from the model, we then resample the values of $x^*$ according to a suitable bootstrap procedure (this is model dependent), to obtain $R$ boostrapped datasets $x^{*r}$ ($r=1,...,R$). For each of these datasets we compute the summary statistics $s^{*r}$ ($r=1,...,R$), and the distances
\[
d_r :=\sqrt{(s^{*r}-s)'\Sigma^{-1}(s^{*r}-s)},\qquad r=1,...,R.
\]
Based on these distances, we calculate how many of the $R$ summaries end up in $\mathcal{D}_j$, and this number is the $n_j$ in \eqref{eq_mu-strat-biased}. Clearly, we have that $n_1+...+n_J=R$.

\paragraph{$\omega_j$ estimation:} here we produce a further independent sample $x^{*'}\sim p(x^*|\theta)$ from the simulator, conditionally to the same $\theta$ used when determining the $n_j$. We obtain $R$ bootstrapped datasets from $x^{*'}$ and the corresponding $R$ summaries. The we compute $R$ distances $d_r$ anew and, as an illustration for the case of three strata over the partition $(0,\delta/2]$, $(\delta/2,\delta]$ and $(\delta,\infty)$, we estimate their probabilities as follows:
\begin{align*}
\hat{\omega}_1 &:= \sum_{r=1}^{R}\mathbb{I}_{\{d_r\leq\delta/2\}}/R, \qquad
\hat{\omega}_2 := \sum_{r=1}^{R}\mathbb{I}_{\{\delta/2< d_r\leq\delta\}}/R,\qquad 
\hat{\omega}_3 :=1-\sum_{j=1}^2\hat{\omega}_j.
\end{align*}
Therefore the $\hat{\omega}_j$ and $n_j$ are obtained independently, so to eliminate the bias that would occur when using the same samples twice. 
The $f$ function in \eqref{eq_mu-strat-biased}, which is the ABC kernel $K_\delta$ in our context, is evaluated only at the summaries simulated when determining the $n_j$.
We now have all the ingredients needed to compute the approximate ABC likelihood $\hat{\hat{\mu}}_{\mathrm{strat}}$ which, following \eqref{eq_mu-strat-biased}, we define as
\begin{equation}
\hat{\hat{\mu}}_{\mathrm{strat}}=\sum_{j=1}^J \biggl\{\frac{\hat{\omega}_j}{n_j} \sum_{i=1}^{n_j} K_\delta(s^{*ij},s)\biggr\}\mathbb{I}_{\{n_j>0,\forall j\}}.\label{eq:abc-strat}
\end{equation}
In the next section we show how the two sets of summaries produced to determine the $n_j$ and the $\hat{\omega}_j$ can be used to produce a likelihood approximation having a smaller variance than $\hat{\hat{\mu}}_{\mathrm{strat}}$, at essentially no additional computational cost.

\subsection{Averaged likelihood by exchanging samples}\label{sec:averaged-likelihood}

Denote with $s^{\mathrm{freq}}$ the collection of $R$ summaries produced to compute the frequencies $n_j$ ($j=1,..,J$) conditionally to some value of $\theta$, and with $s^{\mathrm{prob}}$ the collection of $R$ summaries produced to compute the probabilities $\hat{\omega}_j$ ($j=1,..,J$) conditionally to the same value of $\theta$ as for $s^{\mathrm{freq}}$. We have already shown how to use $s^{\mathrm{freq}}$ and $s^{\mathrm{prob}}$ to enable the construction of $\hat{\hat{\mu}}_{\mathrm{strat}}$. Now set $\hat{\hat{\mu}}_{\mathrm{strat}}^{(1)}:=\hat{\hat{\mu}}_{\mathrm{strat}}$. Since the two sets of summaries are generated independently one of the other, we can construct a second likelihood at zero cost, by exchanging the roles of the two sets of summaries. Namely, this time we use $s^{\mathrm{freq}}$ to obtain a new set of $\hat{\omega}_1,...,\hat{\omega}_J$, except that in this case nothing has to be simulated as we make use of the already available $s^{\mathrm{freq}}$. Similarly, we use $s^{\mathrm{prob}}$ to obtain the $n_1,...,n_J$, again at essentially zero cost. With these new sets of $\hat{\omega}_j$'s and $n_j$'s we construct a second likelihood approximation that we name $\hat{\hat{\mu}}_{\mathrm{strat}}^{(2)}$. Then we average the two likelihoods and obtain 
\begin{equation}
  \bar{\mu}_{\mathrm{strat}}=\frac{\hat{\hat{\mu}}_{\mathrm{strat}}^{(1)}+\hat{\hat{\mu}}_{\mathrm{strat}}^{(2)}}{2}.  \label{eq:avg-like}
\end{equation}
While an explicit expression for $Var(\bar{\mu}_{\mathrm{strat}})$ is unavailable, we show via simulation that $Var(\bar{\mu}_{\mathrm{strat}})<Var(\hat{\hat{\mu}}_{\mathrm{strat}}^{(1)})$ and hence it may be worth considering the averaged likelihood \eqref{eq:avg-like}. See for example section \ref{sec:biased-likelihood} and the results in Supplementary Material. However, whether the averaging approach is appropriate or not it has to be considered on a case-by-case study, as using \eqref{eq:avg-like} doubles the opportunities to obtain some frequency $n_j=0$, hence it comes with a higher rejection rate, see for example section \ref{sec:abc-mcmc} or the samplers in Supplementary Material. On the other hand, in Supplementary Material we show the benefits of computing $\bar{\mu}_{\mathrm{strat}}$ for the case study in section \ref{sec:example-1D-gauss}: when the $n_j>0$ are all positive (both for $\hat{\hat{\mu}}^{(1)}_\mathrm{strat}$ and $\hat{\hat{\mu}}^{(2)}_\mathrm{strat}$) then using $\bar{\mu}_{\mathrm{strat}}$ produces a 50\% reduction in the variance of the likelihood estimation, compared to the one returned via $\hat{\hat{\mu}}_{\mathrm{strat}}^{(1)}:=\hat{\hat{\mu}}_{\mathrm{strat}}$. 

\subsection{Generalized ABC-rejection with stratified Monte Carlo}

In algorithm \ref{abc-gen} we discussed the generalized version of the ABC-rejection algorithm as given in \cite{bornn2017use}.
A version of this algorithm  that uses bootstrapping and stratified Monte Carlo is in algorithm \ref{abc-gen-strat}: notice, the latter includes the possibility to use the ``averaged'' ABC likelihood $\bar{\mu}_{\mathrm{strat}}$ (and in that case we denoted the corresponding normalizing constant as $c_2$ since this is assumed different from the normalizing $c$ pertaining to $\hat{\hat{\mu}}_{\mathrm{strat}}$).
\begin{algorithm}
\scriptsize
\begin{algorithmic}[1]
  \For{$t=1$ to $N$}
    \Repeat 
    \State Generate $\theta^{*} \sim \pi(\theta)$,
   \State 
     generate \textit{once} $x^{i*} \sim p(x| \theta^{*})$ and produce $R$ bootstrapped datasets from $x^{i*}$. Call these $\{x^{r*}\}_{r=1:R}$; 
      \State compute summaries from $\{x^{r*}\}_{r=1:R}$ and obtain the $n_j$, $j=1,...,J$;
     \State Go back to line 3 as soon as some $n_j=0$ otherwise continue;
     \State Generate \textit{once} 
     $x^{i*'} \sim p(x| \theta^{*})$ and produce $R$ bootstrapped datasets from $x^{i*'}$. Call these $\{x^{r*'}\}_{r=1:R}$; 
     \State compute summaries from $\{x^{r*'}\}_{r=1:R}$ and estimate strata probabilities $\omega_j$, $j=1,...,J$;
     \State compute $\hat{\hat{\mu}}_{\mathrm{strat}}$ (or compute ${\bar{\mu}}_{\mathrm{strat}}$);
   \State draw $u \sim \mbox{Uniform}(0,1)$;
    \Until{$u <  \frac{1}{c} \hat{\hat{\mu}}_{\mathrm{strat}}$ (or until $u <  \frac{1}{c_2} \bar{{\mu}}_{\mathrm{strat}}$)}
   \State Set $\theta^{(t)} = \theta^*$\;
  \EndFor
  \end{algorithmic}
\caption{Generalized ABC-rejection with stratified Monte Carlo\label{abc-gen-strat}}
\end{algorithm}
For the determination of the strata and the normalization constant $c$, examples are in the Supplementary Material (regarding $c$, see specifically the supernova example). Generally, the determination of both does not cause an increase in the computational effort in addition to the  preparatory (pilot) runs that are always necessary, as discussed just before section \ref{sec:train-test}.

\section{ABC-MCMC with stratification}\label{sec:abc-mcmc}

For the sake of illustration, here we consider the application of bootstrapping and stratified Monte Carlo into an ABC-MCMC framework (the case of importance sampling ABC is in Supplementary Material). We need to select the generic function $f(\cdot)$ found in \eqref{eq_mu-strat-biased} to be a specific (non-flat) ABC kernel $K_\delta$, for example a Gaussian kernel. In the following, we consider the ABC likelihood as approximated via \eqref{eq:abc-strat}, however we could equivalently consider the one using likelihoods averaging as in \eqref{eq:avg-like}. We write $\hat{\hat{\mu}}^*_\mathrm{strat}\equiv \hat{\hat{\mu}}_\mathrm{strat}(\theta^*)$ and $\hat{\hat{\mu}}^\#_\mathrm{strat}\equiv \hat{\hat{\mu}}_\mathrm{strat}(\theta^\#)$.
A parameter $\theta^*\sim q(\theta^*|\theta^\#)$  is produced via a proposal $g(\cdot)$ and accepted with probability
\begin{equation}
\alpha = \min\biggl\{1,\frac{\hat{\hat{\mu}}_{\mathrm{strat}}^*\pi(\theta^*)}{\hat{\hat{\mu}}_{\mathrm{strat}}^\#\pi(\theta^\#)}\frac{q(\theta^\#|\theta^*)}{q(\theta^*|\theta^\#)}\biggr\}.\label{eq:accept-ratio-stratified}
\end{equation}
For any $\theta^*$ we first check that corresponding summaries generated conditionally on $\theta^*$ have $n_j>0$ for all  $J$ strata: if this is not the case, then $\theta^*$ is immediately rejected, otherwise the $\hat{\omega}_j$ are computed and $\eqref{eq:accept-ratio-stratified}$ is evaluated in a standard Metropolis-Hastings move. Recall that since the $n_j$ and $\hat{\omega}_j$ depend on the parameters, these do not simplify out in the acceptance ratio.
Immediate rejection due to a $n_j$ being zero is the main downside of our approach when $\delta$ is ``very small''. However the benefits of our approach using stratification, as already mentioned in section \ref{sec:stratified-sampling}, come when we use a larger than usual $\delta$, so the issue is mitigated while still obtaining good inference. We now anticipate results discussed in next sections: 
in a pseudomarginal ABC-MCMC (pmABC-MCMC) the likelihood is approximated using several independent samples from the model, as in \eqref{eq:pmABC-lik}. For example, see Figure \ref{fig:gauss-toy}a which is based on $M=500$ independent samples for each value of a scalar $\theta$. When resampling is used to accelerate computations, the variance of the ABC likelihood is reduced, compared to using pmABC-MCMC with $M=1$ without resampling (as often done in practice), as shown in \cite{bootstrapped-sl}. However, a bias is introduced that gives the posterior a larger variance, see Figure \ref{fig:gauss-toy}b which is based on $R=500$ resamples of a single simulated dataset. Instead using stratification (Figure \ref{fig:gauss-toy}c--d) mitigates variance inflation considerably while using a larger threshold $\delta$. In fact, the ABC likelihood with resampling is approximated by the following (unweighted) mean mirroring \eqref{eq:resampled-mc}
\begin{equation}
\hat{\mu}_{\mathrm{res}}=\frac{1}{R}\sum_{r=1}^R K_\delta(s^{*r},s)=\sum_{r=1}^R w_r K_\delta(s^{*r},s),\quad s^{*r}:=S(x^{*r}),\quad x^{*r}\sim\mathrm{res}(x^*)\label{eq:abc-res}
\end{equation}
where weights are constant $w_r=1/R$ (and this is a feature common to standard ABC without resampling). With stratification we have the weighted mean
\begin{equation}
\hat{\hat{\mu}}_{\mathrm{strat}}=\sum_{j=1}^J \biggl\{\hat{w}_j^{\mathrm{strat}} \sum_{i=1}^{n_j} K_\delta(s^{*ij},s)\biggr\}\mathbb{I}_{\{n_j>0,\forall j\}}
\end{equation}
where $\hat{w}_j^{\mathrm{strat}}=\hat{\omega}_j/n_j$.  If we consider the limit case of a single stratum ($J=1$) then $\hat{\omega}_1=1$, $\hat{w}_1^{\mathrm{strat}}=1/{n}_1=1/R$ and  \eqref{eq:abc-strat} reduces to \eqref{eq:abc-res}. 

We can now expand on the considerations we previously expressed about \eqref{eq:biased-expectation}. In fact, if we attempt at taking the expectation of \eqref{eq:abc-strat}, we have (for simplicity, here we assume $\mathbb{I}_{\{n_j>0,\forall j\}}\equiv 1$)
\[
\mathbb{E}(\hat{\hat{\mu}}_{\mathrm{strat}})= \sum_{j=1}^J \mathbb{E}\biggl( \frac{\hat{\omega}_j}{{n}_j(s^*_j)}\sum_{i=1}^{{n}_j}K_\delta(s^{*ij},s)\biggr)\]
where $s_j^*=(s^{*1j},...,s^{*n_jj})$.
Since the summary statistics simulated to produce each $\hat{\omega}_j$  are independent of the $s^{*ij}$  used inside $K_\delta(s^{*ij},s)$ (the latter summaries being independently produced to compute the $n_j$), we have that 
\begin{equation}
\mathbb{E}(\hat{\hat{\mu}}_{\mathrm{strat}})= \sum_{j=1}^J \mathbb{E}\biggl( \frac{\hat{\omega}_j}{{n}_j(s^*_j)}\sum_{i=1}^{{n}_j}K_\delta(s^{*ij},s)\biggr)=
\sum_{j=1}^J \biggl\{\mathbb{E}( \hat{\omega}_j)\mathbb{E}\biggl(\sum_{i=1}^{{n}_j}\frac{K_\delta(s^{*ij},s)}{{n}_j(s^*_j)}\biggr)\biggr\}\label{eq:biased-expectation-2}
\end{equation}
and both expectations are intractable.

Finally, we may plug in the acceptance probability \eqref{eq:accept-ratio-stratified} the estimate $\bar{\mu}_{\mathrm{strat}}$ found in \eqref{eq:avg-like}, in place of $\hat{\hat{\mu}}_\mathrm{strat}$.  However, in this case it is possible that the MCMC acceptance rate will decrease further, as mentioned in section \ref{sec:averaged-likelihood}.

\subsection{Running ABC-MCMC with resampling and stratification}\label{sec:running-abc-mcmc}

We have mentioned the major downside of using stratification is the immediate rejection of a parameter proposal as soon as $n_j=0$ for some stratum. Clearly an improvement is given by introducing a small number of strata (in our examples we always use three strata). Also, the larger the dimension $n_s$ of $s$ the more likely some stratum will be neglected  unless an exaggeratedly large number $R$ of resampled statistics is produced.
\begin{algorithm}
\scriptsize
\caption{ABC-MCMC with resampling (rABC-MCMC)}
\begin{algorithmic}[1]
\State   \textbf{Input:} positive integers $N$ and $R$. Observed summaries $s:=S(x)$ for data $x$. A positive $\delta$ and an ABC kernel $K_\delta(\cdot)$. Fix a starting value $\theta^*$ or generate it from the
prior $\pi(\theta)$. Set $\theta_1:=\theta^*$. A proposal kernel $q(\theta'|\theta)$. Set $l:=1$. \\
\textbf{Output:} $N$ correlated samples from $\pi_\delta(\theta|s)$. \\

\State  \textbf{Initialization:}
\State Given $\theta_1$, generate synthetic data $x^*\sim p(x|\theta_1)$.
\State Generate $R$ datasets $x^{1},...,x^{R}$, each obtained by resampling with replacement from $x^*$. Corresponding summaries are $s^{1},...,s^{R}$. 
\State Compute $\hat{\mu}_\mathrm{res}:=\hat{\mu}_\mathrm{res}(\theta^*)$ as in \eqref{eq:abc-res}. Set $\theta^l:=\theta^*$ and $\hat{\mu}_\mathrm{res}^l:=\hat{\mu}_\mathrm{res}$. Set $l:=l+1$.\\

\State Propose $\theta^*\sim q(\theta|\theta^{l-1})$ and simulate $x^*\sim p(x|\theta^*)$.
\State Generate $R$ datasets $x^{1},...,x^{R}$, each obtained by resampling with replacement from $x^*$. Corresponding summaries are $s^{1},...,s^{R}$. 
\State Compute ${\hat{\mu}}^*_\mathrm{res}:={\hat{\mu}}^*_\mathrm{res}(\theta^*)$ and accept $\theta^*$ with probability
\[
\alpha = \min\biggl\{1,\frac{{\hat{\mu}}_{\mathrm{res}}^*\cdot\pi(\theta^*)}{{\hat{\mu}}_{\mathrm{res}}^{l-1}\cdot\pi(\theta^{l-1})}\times\frac{q(\theta^{l-1}|\theta^*)}{q(\theta^*|\theta^{l-1})}\biggr\}.
\]
If it is accepted, set $\theta^{l}:=\theta^*$ and ${\hat{\mu}}_{\mathrm{res}}^l:={\hat{\mu}}_{\mathrm{res}}^*$, else set $\theta^{l}:=\theta^{l-1}$.
\State Set $l:=l+1$ and go to step 13. 
\State If $l>N$ stop, otherwise go to step 9. 
\end{algorithmic}
\label{alg:rabc-mcmc}
\end{algorithm}

In view of the above, when implementing ABC-MCMC we propose the following strategy: when starting the inference procedure from an initial $\theta^0$, we use an ABC-MCMC with resampling but without stratification for a sufficiently large number of iterations, so that the chain approaches high density regions of the posterior surface. We call this procedure rABC-MCMC (resampling ABC-MCMC) and is exemplified in algorithm \ref{alg:rabc-mcmc}. Once rABC-MCMC has concluded we initialize a resampling procedure with embedded stratification using \eqref{eq:abc-strat} for a number of additional iterations, to obtain a refined chain to be used for reporting results. We call this second stage rsABC-MCMC (resampling ABC-MCMC with stratification). The advantage of starting the simulation with rABC-MCMC is that it is fast and empirically is shown to be able to locate the mode of the ABC posterior $\pi_\delta(\theta|s)$. In Supplementary Material we give a suggestion for how we let the threshold $\delta$ decrease (this is not central for our work), so when rsABC-MCMC starts it uses the $\delta$ returned by the last iteration of rABC-MCMC, as well as a tuned scaling matrix $\Sigma$. 
Basically, we use rsABC-MCMC to refine the inference produced by the over-dispersed rABC-MCMC chain. We do not strictly need to further reduce $\delta$ during the stratification stage. Using these settings, rsABC-MCMC is illustrated in algorithm \ref{alg:rsabc-mcmc}. Instead, when rsABC-MCMC uses an ABC likelihood that is approximated via \eqref{eq:avg-like}, then we call the resulting algorithm xrsABC-MCMC, where the ``x'' stands for the ``e\texttt{x}change'' of the  summaries role when computing the second likelihood in $\bar{\mu}_\mathrm{strat}$.
\begin{algorithm}
\scriptsize
\caption{ABC-MCMC with resampling and stratification (rsABC-MCMC)}
\begin{algorithmic}[1]
\State   \textbf{Input:} positive integers $N$, and $R$. Observed summaries $s:=S(x)$ for data $x$. A positive $\delta$ possibly inherited from algorithm \ref{alg:rabc-mcmc}, an ABC kernel $K_\delta(\cdot)$ and $J$ strata $\mathcal{D}_j$. Set a starting value $\theta^*$ possibly obtained from algorithm \ref{alg:rabc-mcmc}. Set $\theta^1:=\theta^*$. A proposal kernel $q(\theta'|\theta)$ possibly inherited from algorithm \ref{alg:rabc-mcmc}. Set $l:=1$. \\
\textbf{Output:} $N$ correlated samples from $\pi_\delta(\theta|s)$. \\

\State  \textbf{Initialization:}
\State Given $\theta_1$, generate synthetic data $x^*\sim p(x|\theta^1)$.
\State Generate $R$ datasets $x^{1},...,x^{R}$, each obtained by resampling with replacement from $x^*$. Corresponding summaries are $s^{1},...,s^{R}$. Obtain $n_1,...,n_J$ from the strata  $\mathcal{D}_1,...,\mathcal{D}_J$. 
\State If some $n_j=0$ is found go back to step 5 (we cannot continue otherwise without a nonzero $\hat{\hat{\mu}}^1_\mathrm{strat}$), otherwise go to next step.
\State Generate another independent dataset $x^{*'}\sim p(x|\theta^1)$.  Generate $R$  datasets $x^{1},...,x^{R}$, each obtained by resampling with replacement from $x^{*'}$. Obtain summaries $s^{1},...,s^{R}$ anew to produce $\hat{\omega}_1,...,\hat{\omega}_J$.
\State Compute $\hat{\hat{\mu}}^1_\mathrm{strat}$ as in \eqref{eq:abc-strat}. Set $l:=2$.
\State
\State Propose $\theta^*\sim q(\theta|\theta^{l-1})$, simulate $x^*\sim p(x|\theta^*)$ and compute the $n_j$ as above.
\State If some $n_j=0$ reject immediately: set $\theta^{l}:=\theta^{l-1}$ and go to step 15, otherwise go to next step. 
\State Simulate $x^{*'}\sim p(x|\theta^*)$, resample $R$ datasets, obtain the summaries and compute the $\hat{\omega}_1,...,\hat{\omega}_J$ as usual.
\State Compute $\hat{\hat{\mu}}^*_\mathrm{strat}:=\hat{\hat{\mu}}^*_\mathrm{strat}(\theta^*)$ as in \eqref{eq:abc-strat} and accept $\theta^*$ with probability
\[
\alpha = \min\biggl\{1,\frac{\hat{\hat{\mu}}_{\mathrm{strat}}^*\cdot\pi(\theta^*)}{\hat{\hat{\mu}}_{\mathrm{strat}}^{l-1}\cdot\pi(\theta^{l-1})}\times\frac{q(\theta^{l-1}|\theta^*)}{q(\theta^*|\theta^{l-1})}\biggr\}.
\]
If it is accepted, set $\theta^{l}:=\theta^*$ and $\hat{\hat{\mu}}_{\mathrm{strat}}^l:=\hat{\hat{\mu}}_{\mathrm{strat}}^*$, else set $\theta^{l}:=\theta^{l-1}$.
\State Set $l:=l+1$ and go to step 16. 
\State If $l>N$ stop, otherwise go to step 11.
\end{algorithmic}
\label{alg:rsabc-mcmc}
\end{algorithm}

\section{Case studies}\label{sec:case-studies}

Supporting code for the case studies is available at \url{https://github.com/umbertopicchini/stratABC}. For all ABC approaches $K_\delta$ is always a Gaussian kernel.  

\subsection{A Gaussian toy-model}\label{sec:example-1D-gauss}

\begin{figure}
    \centering
    \begin{subfigure}[b]{0.18\textwidth}
        \includegraphics[width=\textwidth]{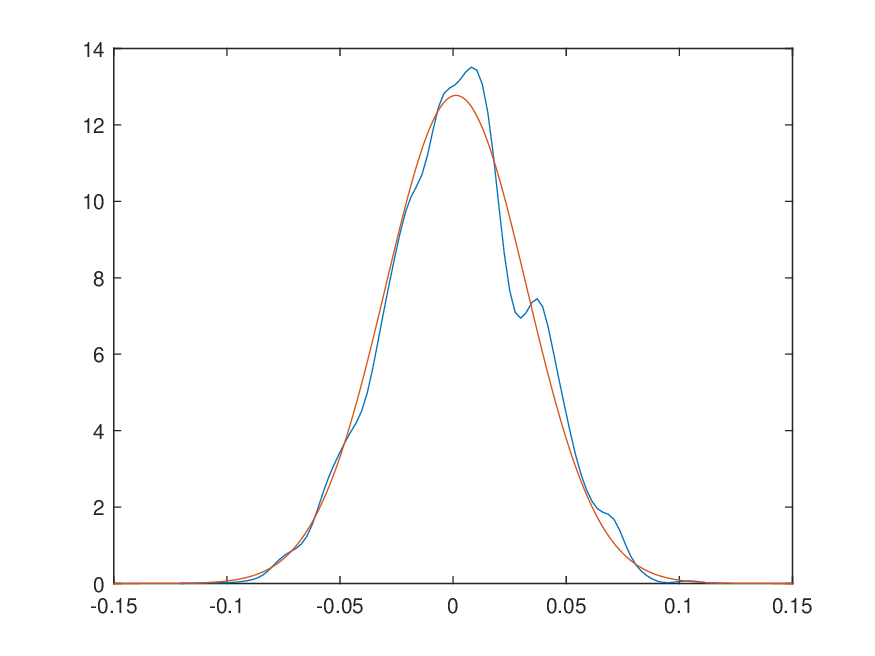}
        \caption{\footnotesize{pmABC, $\delta=3\cdot 10^{-5}$, $M=500$}}
    \end{subfigure}
    ~ 
    \begin{subfigure}[b]{0.18\textwidth}
        \includegraphics[width=\textwidth]{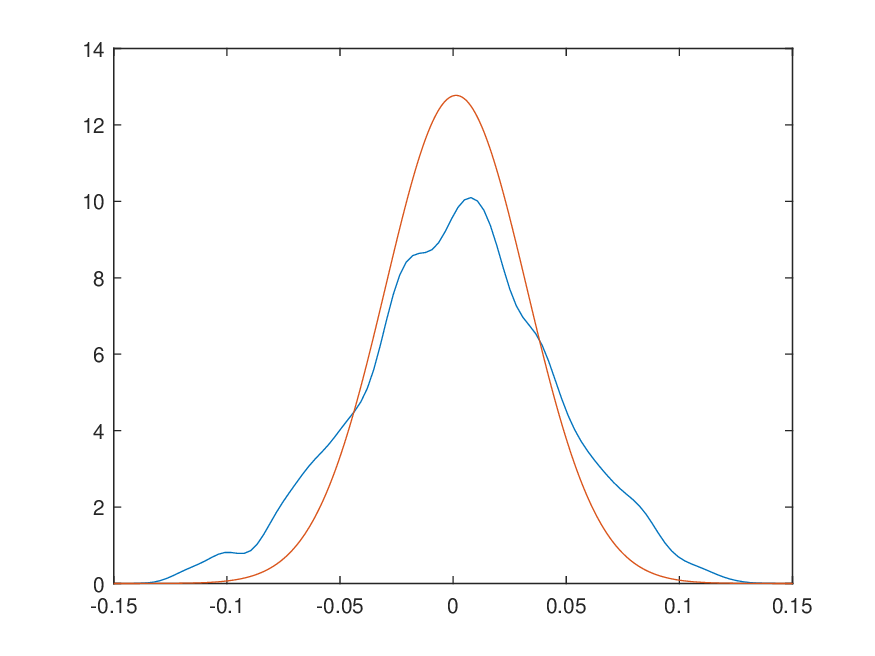}
        \caption{\footnotesize{rABC, $\delta=3\cdot 10^{-5}$, $R=500$}}
    \end{subfigure}
    ~ 
    \begin{subfigure}[b]{0.18\textwidth}
        \includegraphics[width=\textwidth]{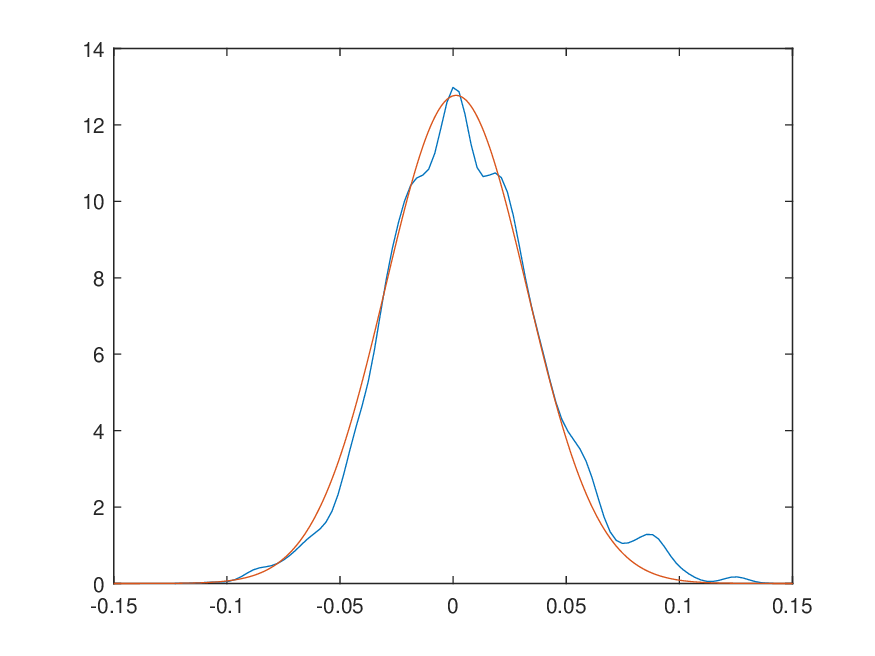}
        \caption{\footnotesize{rsABC, $\delta=3\cdot 10^{-4}$, $R=500$}}
    \end{subfigure}
    ~
    \begin{subfigure}[b]{0.18\textwidth}
        \includegraphics[width=\textwidth]{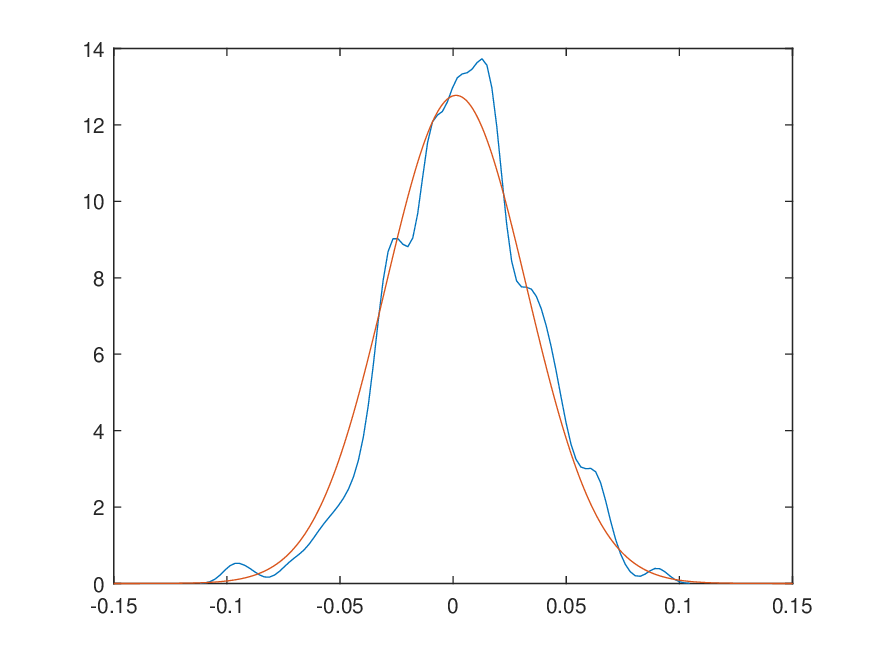}
        \caption{\footnotesize{xrsABC, $\delta=3\cdot 10^{-4}$, $R=500$}}
    \end{subfigure}
    ~
    \begin{subfigure}[b]{0.18\textwidth}
        \includegraphics[width=\textwidth]{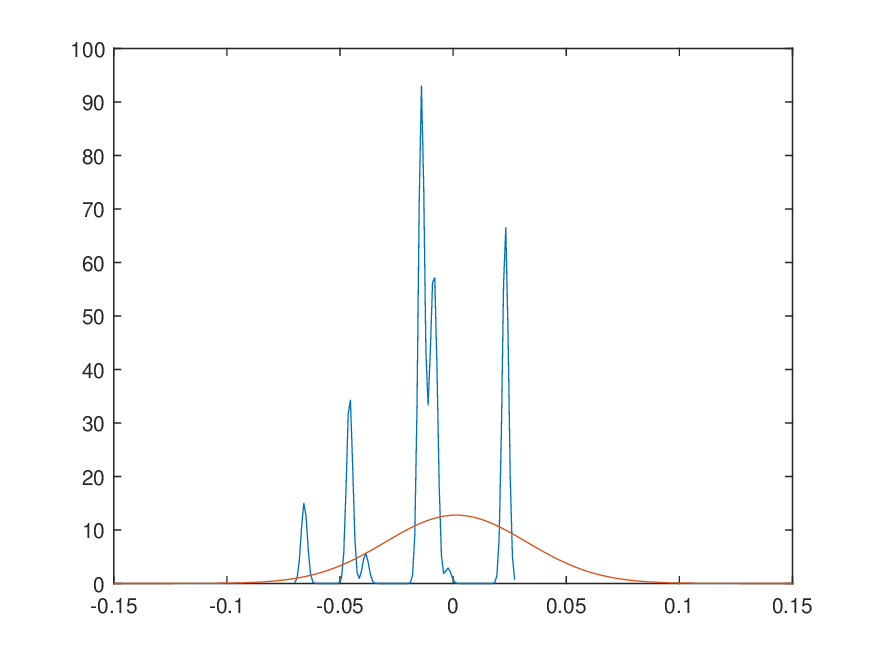}
        \caption{\footnotesize{pmABC, $\delta=3\cdot 10^{-5}$, $M=2$}}
    \end{subfigure}
    \caption{\footnotesize{Gaussian model: posteriors for $\theta$. The exact posterior is in red. Approximate posteriors are in blue.}}
    \label{fig:gauss-toy}
\end{figure}

Here we consider a trivial but illustrative example of the consequences of using stratified Monte Carlo sampling within ABC.
Data $x$ is a sample of $n_\mathrm{obs}=1,000$ iid realizations from a standard Gaussian. We conduct Bayesian inference for the population mean $\theta$, and assume that our data-generating model for a single entry of $x$ is $\mathcal{N}(\theta,1)$, i.e. the variance is known but the mean is unknown. We assume Gaussian priors (these are conjugate) so that exact Bayesian inference is possible. Specifically, we set $\pi(\theta)\sim \mathcal{N}(m_0,\sigma_0^2)$ with hyperparameters $m_0=0.1$ and $\sigma_0=0.2$.
As an illustration, we consider sampling via ABC-MCMC, however in this example we are not interested in the convergence of ABC-MCMC algorithms for values of $\theta$ starting far away from the truth, and we initialize $\theta$ at its ground-truth value $\theta=0$. For this problem we also have a sufficient summary statistic for $\theta$, which is the sample mean of the data. Therefore the value of $\delta>0$ is the only source of approximation (besides the Monte Carlo error due to finite sampling).  Results in Figure \ref{fig:gauss-toy} are based on 9,000 iterations following 1,000 burnin iterations. Parameter proposals are always generated using Gaussian random walk with (fixed) standard deviation of 0.1. 
We now describe four ABC procedures.
\textbf{pmABC-MCMC:} here we use \eqref{eq:pmABC-lik} into a Metropolis-Hastings algorithm with $M=500$ and $\delta=3\times 10^{-5}$. That is, for each $\theta^*$ we produce $M$ independent datasets $x^{*r}$ ($r=1,...,M$), each having $n_\mathrm{obs}$ entries simulated iid from $\mathcal{N}(\theta^*,1)$. Then the summaries $s^{*r}$ of each $x^{*r}$ are taken, and the ABC likelihood is approximated by averaging the $M$ ABC kernels. The resulting posterior density is in Figure \ref{fig:gauss-toy}(a). The acceptance rate is around 12\%.
\textbf{rABC-MCMC:} here rABC-MCMC stands for ABC-MCMC with resampling (but no stratification) and is similar to pmABC-MCMC except for the use of data resampling. For each new $\theta^*$, we obtain a single realization $x^*$ of size $n_\mathrm{obs}$ from the model simulator, which we bootstrap (non-parametrically) to obtain $R=500$ vectors $x^{*r}$, each having size $n_\mathrm{obs}$, $r=1,...,R$. For each vector we compute summaries $s^{*r}$ and then \eqref{eq:abc-res}.
The resulting posterior obtained with $\delta=3\times 10^{-5}$ is in Figure \ref{fig:gauss-toy}(b). We notice the detrimental effect of resampling, inflating the posterior variability considerably. However the mean is correctly estimated. The acceptance rate is around 11\%.
\textbf{rsABC-MCMC:} here rsABC-MCMC stands for ABC-MCMC with resampling and stratification. Here we use $\delta_{rs}=3\times 10^{-4}$, that is a threshold ten times larger than before. 
We use three strata defined as $\mathcal{D}_1=\{s^* \text{ s.t. } \sqrt{\sum_r(s^{*r}-s)^2}\in (0,\delta_{rs}/2)\}$, $\mathcal{D}_2=\{s^* \text{ s.t. } \sqrt{\sum_r(s^{*r}-s)^2}\in (\delta_{rs}/2,\delta_{rs})\}$ and $\mathcal{D}_3=\{s^* \text{ s.t. } \sqrt{\sum_r(s^{*r}-s)^2}\in (\delta_{rs},\infty)\}$ and use $R=500$. Results are in Figure \ref{fig:gauss-toy}(c) . We clearly note the improvement over rABC-MCMC and obtain results similar to pmABC-MCMC. In addition, the acceptance rate is about 70\%, much larger than for the other schemes. As previously mentioned, strata do not have in general to depend on $\delta_{rs}$. Other case studies also use strata determined from prior-predictive simulations.
\textbf{xrsABC-MCMC:} here xrsABC-MCMC is the same as rsABC-MCMC, except that we average the two stratified likelihoods, as described in section \ref{sec:averaged-likelihood}. The inference appears slightly worse than rsABC-MCMC and this is because of the lower mixing, caused by the lower acceptance rate of 6\% which is induced by the higher opportunity to neglect a stratum for this scheme. However, as detailed in next section, this schemes produces a likelihood with a much smaller variance compared to rsABC-MCMC. 

Finally, since \textit{accepted} draws from (x)rsABC-MCMC result from resampled data based on two independent  model simulations (one for the $n_j$ and one for the $\omega_j$), it is interesting to see what happens when pmABC-MCMC  is run with $M=2$. This is illustrated in Figure \ref{fig:gauss-toy}(e), resulting from an acceptance rate of 1\%. In this case the ABC likelihood has a very large variance, thus producing sticky patterns in the chain. While observing such acceptance rate is common in ABC studies with $M=1$, this implies that very long chains are required to explore the posterior adequately, which is not always a possibility with expensive model simulator. See in particular section \ref{sec:lv}.

\subsubsection{Likelihood estimation}\label{sec:biased-likelihood}

\begin{figure}
\centering
\begin{subfigure}[b]{0.35\textwidth}
\includegraphics[width=\textwidth]{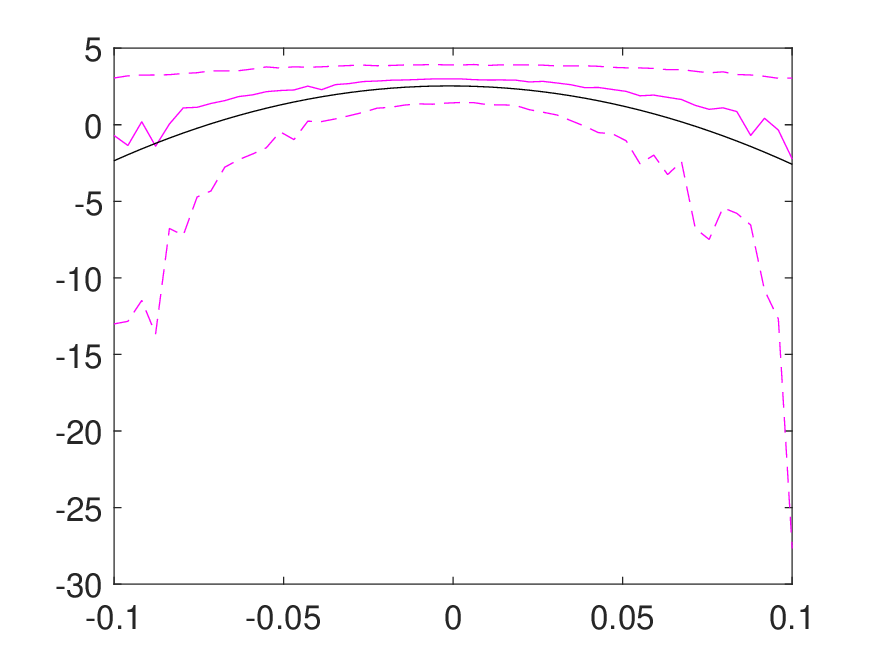}
\caption{\footnotesize{exact loglikelihood (black) and ABC loglikelihood via resampling with stratification (magenta).}}\label{fig:gauss-exactlik-rsabclik}
\end{subfigure}
\qquad
\begin{subfigure}[b]{0.35\textwidth}
\includegraphics[width=\textwidth]{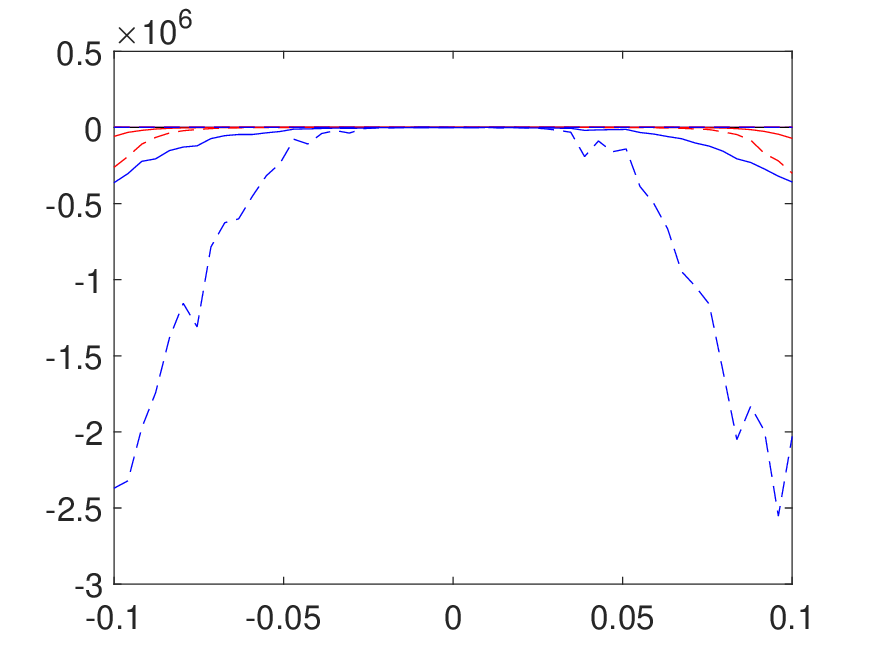}
\caption{\footnotesize{ABC loglikelihood estimated via standard ABC (in red) and via rABC (blue).}}\label{fig:gauss-three-likelihoods}
\end{subfigure}
\caption{\footnotesize{1D Gauss model: exact and approximate loglikelihoods of $s$. Solid lines are means across 1000 estimation attempts. Dashed lines are 2.5 and 97.5 percentiles. Notice the different order of magnitude for the y-axes.}}
\end{figure}

We now consider the estimation of the ABC likelihood function, regardless of Bayesian inference. We apply the four likelihood estimation strategies employed in section \ref{sec:example-1D-gauss}.
We compare: (i) standard Monte Carlo ABC (here denoted ABC) which uses $M$ independent samples, i.e. equation \eqref{eq:pmABC-lik}; (ii) resampling, as in equation \eqref{eq:abc-res} (denoted rABC), (iii) resampling with stratified sampling (denoted rsABC) as in equation \eqref{eq:abc-strat}, and finally (iv) in Supplementary Material we report the variance reduction implied by averaging likelihoods by ``exchanging samples'', as described in section \ref{sec:averaged-likelihood}. We use the same data previously considered and the same values $M=500$, $R=500$ and $\delta$ as considered in the previous experiments, that is $\delta=3\cdot 10^{-5}$ for standard ABC and rABC and the ten times larger $\delta=3\cdot 10^{-4}$ for rsABC. We compare the likelihoods at fifty equally spaced values taken in the interval $\theta\in [-0.1,0.1]$.  For each of the fifty considered values of $\theta$, the several approximate likelihoods are computed independently for 1000 times, using the same settings as previously introduced. For ease of visual comparison, results are given for the corresponding loglikelihoods.
The only difference with the previous MCMC inference is that, when using stratification for this specific study, we need to impose that $n_j>0$ for all $j$ and all estimation attempts. That is in this specific section (and only here) with rsABC we keep simulating until we obtain realizations that satisfy such constraints (or otherwise we would not produce loglikelihoods having finite values, hence these would be useless for display). Such restriction is not necessary (nor implemented) when Bayesian inference is conducted with our methods. 
 We first compare in Figure  \ref{fig:gauss-exactlik-rsabclik} the exact loglikelihood of the summary statistic with the one approximated via rsABC. Then Figure  \ref{fig:gauss-three-likelihoods}  displays the very high variance of rABC, in the order of $10^6$ in the tails \textit{on the log-scale}, and in the order of $10^5$ in the tails of standard ABC.
 Therefore methods display a much higher variation compared to the stratified approximation (but remember the latter is computed only when all $n_j>0$ and this requires many attempts to be achieved). In Supplementary Material we show a zoomed-in version into the central part of Figure \ref{fig:gauss-three-likelihoods}, this revealing that both standard ABC and rABC returned a much more biased likelihood approximation, compared to rsABC. Notice the average of the rABC estimation is not even appearing in such plot since its bias is so large.
In Supplementary Material we show the benefits of computing $\bar{\mu}_{\mathrm{strat}}$ (the likelihood obtained by ``exchanging samples''): when the $n_j>0$ are all positive (both for $\hat{\hat{\mu}}^{(1)}_\mathrm{strat}$ and $\hat{\hat{\mu}}^{(2)}_\mathrm{strat}$) then using $\bar{\mu}_{\mathrm{strat}}$ produces a 50\% reduction in the variance of the likelihood estimation, compared to the one returned via rsABC. 

\subsection{Supernova cosmological model}\label{sec:supernova}

We present an astronomical example initially studied in \cite{jennings2017astroabc}, and recently reconsidered in \cite{picchini2020adaptive}, and compare many different sampling approaches for simulation-based inference. We use the same data  and setup as in \cite{picchini2020adaptive}, to which we refer the reader for full details. Briefly, supernovae redshifts with range $z \in [0.5, 1.0]$ are simulated and then binned into $20$ redshift bins (simulation details are in Supplementary Material). The model that describes the distance modulus as a function of redshift $z$, known in the astronomical literature as Friedmann--Robertson--Model \citep{condon2018lambdacdm}, is:  

\begin{equation}
\mu_{i}(z_{i};\Omega_{m},\Omega_{\Lambda},\Omega_{k},w_{0}, h_0) \propto 5\log_{10}\biggl(\frac{c(1+z_i)}{h_{0}}\biggr)\int_{0}^{z_i}\frac{dz'}{E(z')},
\label{eq:supernovaFM}
\end{equation}
where $E(z) = \sqrt{\Omega_{m} (1+z)^{3} + \Omega_{k} (1+z)^{2} + \Omega_{\Lambda}e^{3\int_{0}^{z} dln(1+z')[1+w(z')]}}$.

The cosmological parameters involved in \eqref{eq:supernovaFM} are five. The first three parameters are the matter density of the universe, $\Omega_m$, the dark energy density of the universe, $\Omega_{\Lambda}$ and the radiation and relic neutrinos, $\Omega_{k}$. The final two parameters are, respectively, the present value of the dark energy equation, $w_0$, and the Hubble constant, $h_0$. A number of parameter constraints and simplifications are then considered in \cite{picchini2020adaptive} which we do not describe here. For the present example, ground-truth parameters are set as follows: $\Omega_m = 0.3$, $\Omega_k = 0$, $w_0 = -1.0$ and $h_0 = 0.7$.
Parameters $h_0$ and $\Omega_k$ are assumed known. Similarly to \cite{jennings2017astroabc} and \cite{picchini2020adaptive},  we aim at inferring the cosmological parameters $\theta=(\Omega_m,w_0)$. In \cite{picchini2020adaptive}  several approximate methods are compared, namely the adaptive ABC–PMC (aABC–PMC) of \cite{simola2021adaptive}, Bayesian optimization via BOLFI \citep{gutmann2016bayesian}, Bayesian synthetic likelihoods with shrinkage-robustified covariance (sBSL) from \cite{nott2019bayesian}, and correlated synthetic likelihoods using the guided and adaptive MCMC sampler of \cite{picchini2020adaptive}. We do not further detail the last four methods and refer the interested reader to \cite{picchini2020adaptive}.
We consider $s=(\mu_1,...,\mu_{20})$ as summary statistic. Notice, when data are simulated as illustrated above, $s$ is a trivial summary statistic, in that $(\mu_1,...,\mu_{20})$ is the data itself (since both the $u_j$ and the $z_i$ do not depend on $\theta$).  For all experiments, we set priors $\Omega_m \sim Beta(3, 3)$, since $\Omega_m$ must be in $(0,1)$, and $w_0 \sim \mathcal{N}(-0.5, 0.5^2)$. 

We consider: the  (generalized) ABC-rejection and our version enhanced with stratified Monte Carlo, that is algorithm \ref{abc-gen} and algorithm \ref{abc-gen-strat} respectively; ABC importance sampling (ABC-IS) and our stratified Monte Carlo version (sABC-IS), which are detailed in the Supplementary Material.
The generalized ABC-rejection (gABC-rej), our stratified Monte Carlo version (sgABC-rej), ABC-IS and our sABC-IS were all set to produce 1,000 posterior samples with $\delta=0.15$. For sgABC-rej and sABC-IS, the construction of the strata is discussed in the Supplementary Material. 
For sBSL, ACSL and BOLFI, we obtained 10,000 posterior samples (after burnin), which were thinned to finally obtain 1,000 posterior samples (to match the number of particles produced via aABC-PMC, which is also 1,000). We also report the effective sample size (ESS) for the 1,000 thinned MCMC draws. The ESS is obtained from the \texttt{coda} R package \citep{coda} and we report as minESS the ESS corresponding to the worst performing chain among $\Omega_m$ and $w_0$. Our main interest here is in the comparison between gABC-rej, sgABC-rej, ABC-IS and sABC-IS, while results for the other methods are merely reported for the interested reader, who is invited to refer to \cite{picchini2020adaptive}. 
For the methods of interest, we also consider wall-clock runtime. It is not possible to perform a precise comparison in terms of runtime between methods using stratified Monte Carlo and  gABC-rej or ABC-IS. This is because while gABC-rej or ABC-IS use $M$ model simulations at each proposed parameter, instead for sgABC-rej and sABC-IS this is not always the case: in fact, when some stratum is neglected ($n_j=0$) the proposal is immediately rejected (and only one model simulation is performed, hence $M=1$), but when  $n_j>0$ for all $j$ at that proposed parameter, then a second model simulation is performed to estimate the $\omega_j$ (hence $M=2$). Therefore, the best we can do is to present results for gABC-rej and ABC-IS for both cases $M=1$ and $M=2$. 

ABC-rejection algorithms produce independent samples, and therefore we do not report their ESS values (all samples are ``effective''). We report instead the wall-clock run-times and the total number of proposed parameters that were necessary to produce 1,000 accepted posterior draws.  We compare gABC-rej, sgABC-rej, ABC-IS and sIS-AB to  aABC-PMC. Of course the rejection-based ABC methods are intrinsically wasteful, this is expected, but our sgABC-rej is about 3.2 times more efficient in producing acceptances than gABC-rej (compare the number of proposals produced to accept 1,000 of those). The credible intervals of $w_0$ from sgABC-rej are closer to the aABC-PMC ones.
For the importance sampling versions, clearly the algorithms are less wasteful, and while we notice similar inference between ABC-IS and sABC-IS, we observe that with stratified Monte Carlo we require about half the number of proposals to obtain 1,000 acceptances. In terms of running time, the overhead induced by data resampling can certainly be lowered if the bootstrapping part of our code was written in some efficient language such as C/C++ (something we didn't do), which means that further acceleration can be achieved.

\begin{table}[ht]
\tiny
\begin{center}
\resizebox{\columnwidth}{!}{%
\begin{tabular}{cccccc}
\hline 
 & aABC--PMC & sBSL  & ACSL & BOLFI\\
\hline 
$\Omega_m$ &  0.31 (0.071, 0.540)  & 0.313 (0.136, 0.474)  & 0.317 (0.129, 0.488) & 0.289 (0.0765, 0.467)\\ 
$w_0$ & -1.05 (-1.95, -0.520)  &  -1.014 (-1.517 -0.580) & 1.047 (-1.502, -0.563) & -0.99 (-1.540, -0.545)\\
minESS & -- & 301 & 681 & 831\\
\hline
 & gABC-rej (M=1) & gABC-rej (M=2) & sgABC-rej &  \\
 \hline
 $\Omega_m$ & 0.323 [0.103, 0.537] & 0.327 [0.117, 0.547] & 0.342 [0.106, 0.573] &  \\
$w_0$ & -0.993 [-1.548, -0.567] & -1.002 [-1.629, -0.567] & -0.984 [-1.732, -0.520] &  \\
runtime (min) & 6.6 & 15.6 & \textbf{3.5} \\
\# proposals & $4.1\cdot 10^4$ & $4.1\cdot 10^4$ & $\mathbf{1.28\cdot 10^4}$ \\
ESS & -- & -- & --  \\
\hline
 &ABC-IS (M=1, $\delta=0.15$) & ABC-IS (M=2, $\delta=0.15$)  &  sABC-IS ($\delta=0.75$) \\
 \hline
 $\Omega_m$ & 0.323 [0.105, 0.525] & 0.329 [0.115, 0.533] & 0.340 [0.120, 0.589]\\
$w_0$ & -1.020 [-1.644, -0.595] & -1.011 [-1.698, -0.562] &  -0.979 [-1.652, -0.521]\\
runtime (min) & 2.7 & 5.7 & \textbf{2.0}\\
\# proposals & $1.7\cdot 10^4$ & $1.8\cdot 10^4$ & $\mathbf{0.91\cdot 10^4}$\\
ESS & \textbf{767} & 713 & 700\\
\hline
\end{tabular}
}
\caption{\footnotesize{Supernova model: posterior means (95\% HPD interval) from several methods. Ground-truth parameters are $(\Omega_m,w_0)=( 0.3,-1)$. We also report the wallclock runtime (minutes), the total number of attempted proposals required to obtain 1,000 acceptances (lower is better), and the effective sample size of the latter (larger is better).}}
\label{table.results.mh}
\end{center}
\end{table}

\subsection{Ising model}

\begin{figure}
    \centering
    \includegraphics[scale=0.5]{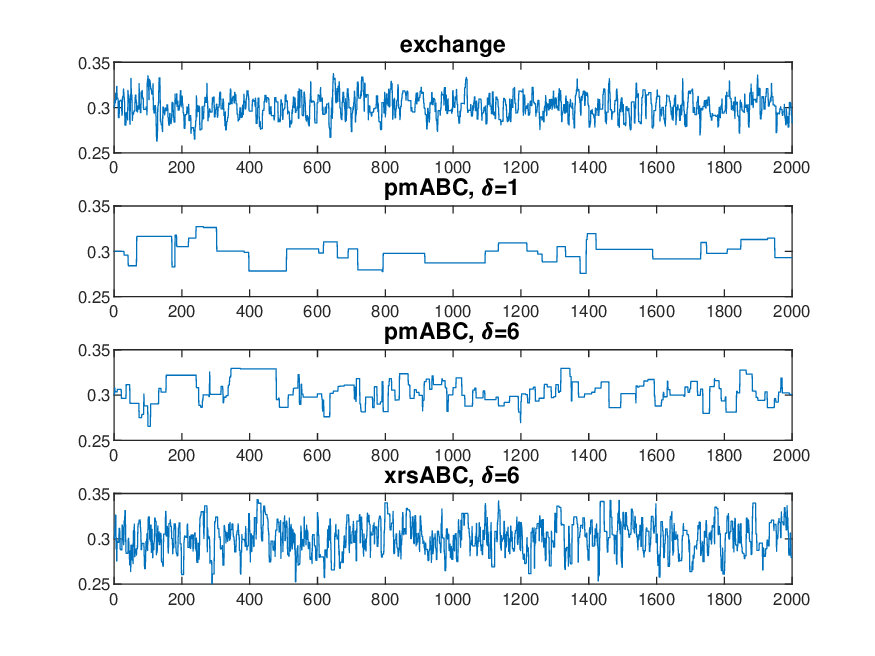}
    \caption{\footnotesize{Ising model: 2,000 iterations of the exchange algorithm (top), pmABC-MCMC with $M=2$ and $\delta=1$ (second row), pmABC-MCMC with $M=2$ and $\delta=6$ (third row) and xrsABC-MCMC with $R=500$ and $\delta=6$ (bottom).}}
    \label{fig:ising_traces}
\end{figure}

The Ising model is a Markov random field model for a vector of binary
variables $x=\left(x_{k}\right)_{k=1}^{n_\mathrm{obs}}$, each of which takes a
value in $\left\{ -1,1\right\} $. Each variable $x_{k}$ has a set
of neighbouring variables $\aleph_{k}\left(x_{k}\right)$, and the
joint distribution over $x$ is given by
\[
f\left(x\mid\theta\right)=\frac{\exp\left(\theta S\left(x\right)\right)}{Z\left(\theta\right)},
\]
where $S\left(x\right)=\sum_{k=1}^{n_\mathrm{obs}}\sum_{x_{\aleph}\in\aleph\left(x_{k}\right)}x_{k}x_{\aleph}$
and $Z\left(\theta\right)$ (the ``partition function'') is the
sum of the numerator over all possible configurations of $x$ and
is thus usually too computationally expensive to evaluate pointwise
at $\theta$. In this paper we consider the well-studied case when
$\aleph_{k}$ has the $x$ variables arranged in a 2-dimensional grid
(using toroidal boundary conditions). Exact Bayesian inference of the parameter
$\theta$ can be performed using the ``exchange algorithm'' introduced in \cite{Murray2006}.
In this paper we study simulated data generated for a $100\times100$
grid (so that $n_\mathrm{obs}=10^{4}$), using the parameter $\theta=0.3$.
This is a computationally expensive model, so we ran the exchange algorithm for 2,000 iterations 
on this data and compared with results from different ABC-MCMC algorithms, using the same number of iterations. A uniform prior between 0 and 3 was used for $\theta$. The algorithms were run 40 times each, and in all cases
were initialised at $\theta=0.3$ (so that we can compare their efficiency
without requiring a burn in). To simulate from the model, a Gibbs
sampler with 50 sweeps was used, with the final point being taken
as the simulated value for the ABC. The statistic $S(\cdot)$ (defined
above) was used in the ABC, and employed a Gaussian ABC kernel
(noting that we do so even though $S\left(\cdot\right)$ only takes
discrete values). When running pmABC-MCMC we use $M=2$ since with stratified ABC we also simulate twice from the model when every $n_j$ is positive. We considered two values for the ABC threshold, namely $\delta=1$ and $\delta=6$. The first value was of interest because, when running pmABC-MCMC with $M=2$ and $\delta=1$ the observed acceptance rate was about 1\%, which is typically a good compromise in ABC studies between inference accuracy and computational effort. However, for such a limited number of iterations, $\delta=1$ produces a very poorly mixing chain. An example from one of the forty runs is in Figure \ref{fig:ising_traces} (second row), and all other runs show similar behaviours.

\begin{figure}
    \centering
    \begin{subfigure}[b]{0.38\textwidth}
    \includegraphics[width=\textwidth]{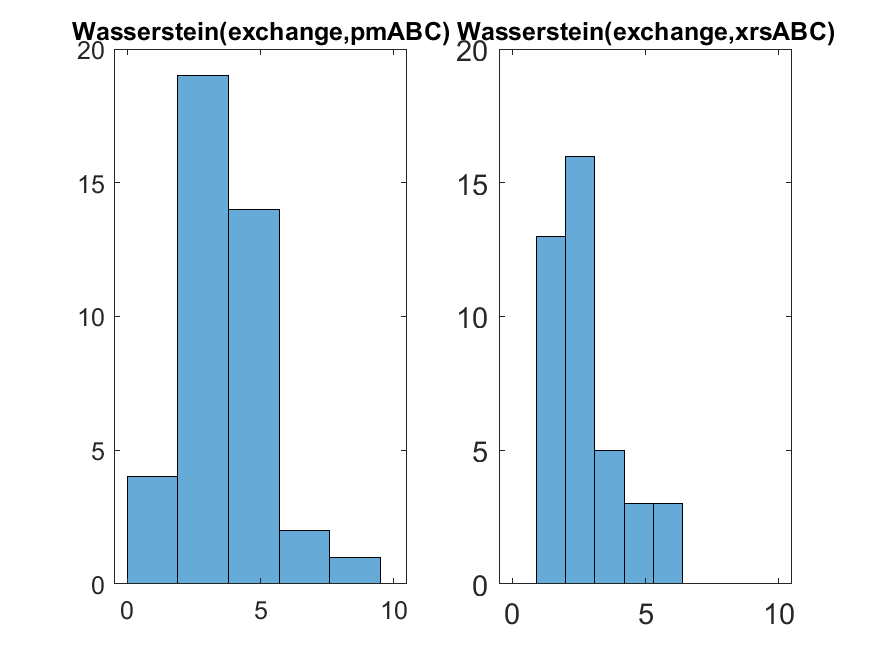}
    \caption{}
    \label{fig:wasserstein_histograms}
    \end{subfigure}
    \begin{subfigure}[b]{0.32\textwidth}
        \includegraphics[width=\textwidth]{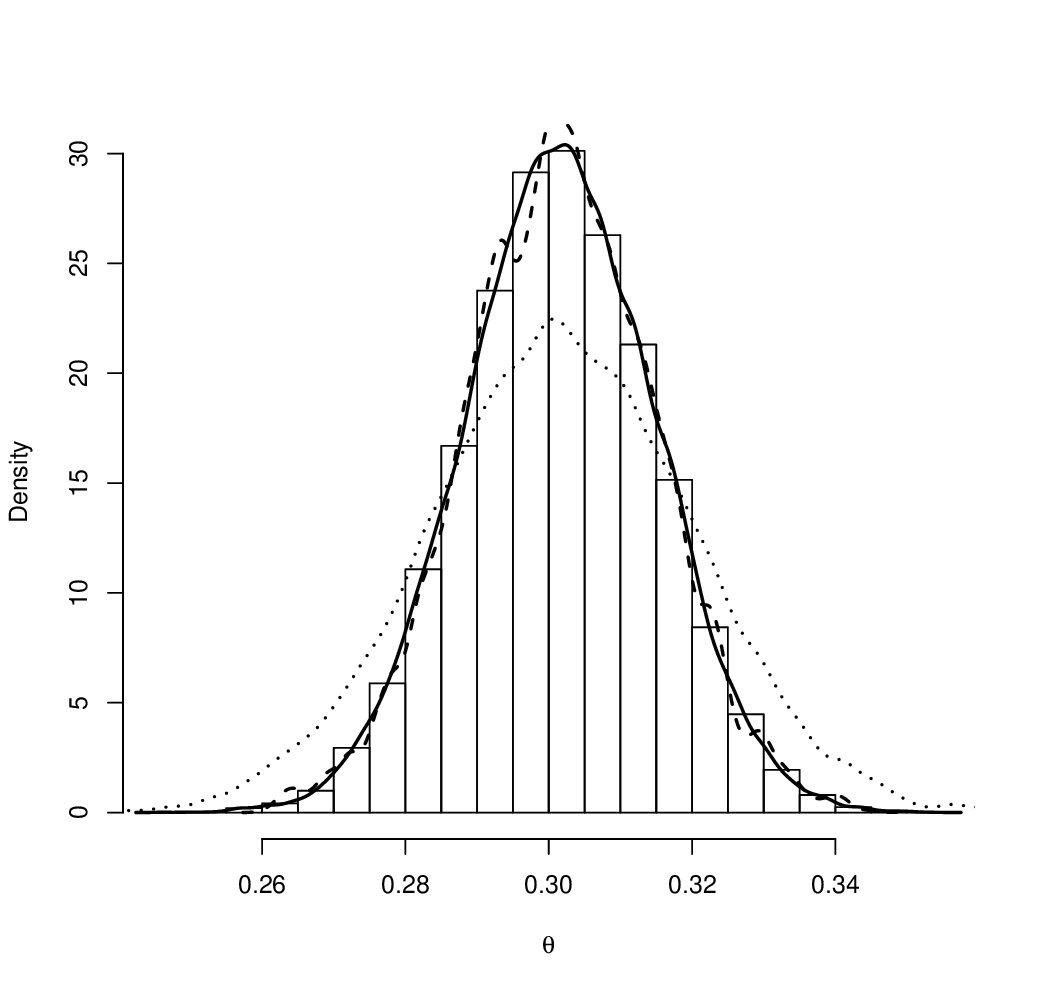}
    \caption{}
    \label{fig:ising_histograms}
    \end{subfigure}
    \caption{\footnotesize{Ising model: (a) histograms of forty Wasserstein distances between the posteriors from the exchange algorithm and pmABC (left) and between the posteriors from the exchange algorithm and xrsABC (right). (b) histogram of 80,000 draws from the exchange algorithm and corresponding kernel smoothing line (solid). We also have the kernel smoothing line for the pmABC (dashed) and xrsABC posteriors (dotted).}}
\end{figure}

We then increase the threshold to $\delta=6$ and produce results for both pmABC-MCMC ($M=2$) and for xrsABC-MCMC ($R=500$). For the latter, resampling was performed using the block bootstrap as in
\citet{bootstrapped-sl} with a block size of $20\times20$.
For xrsABC-MCMC with $\delta=6$ we considered $\mathcal{D}_1= (0,\delta/2)$, $\mathcal{D}_2=[\delta/2,\delta)$, $\mathcal{D}_3=[\delta,\infty)$. The illustrative plot in Figure \ref{fig:ising_traces} shows that pmABC-MCMC has poor mixing  even with $\delta=6$. For example, a measure of sampling efficiency is the integrated autocorrelation time (IAT), which we compute via the \texttt{LaplacesDemon} R package \citep{laplaces} and gives the number of iterations required to obtain an independent sample (hence the smaller the IAT the better). We compute the IAT for each of the forty repetitions and report their sample averages. The mean IAT for the exchange algorithm is 9.4, while for pmABC ($\delta=6$) is 38.6 and for xrsABC ($\delta=6$) is 10.5. If we were to compute posterior quantiles, it would appear that the results of pmABC-MCMC (with $\delta=6$) are, after all, quite accurate, but that would be an illusion. As an example, for the illustrative run in Figure \ref{fig:ising_traces}, the gold-standard results for $\theta$ provided by the exchange algorithm are (posterior mean and 95\% intervals) 0.301 [0.277,0.326], for pmABC-MCMC ($\delta=6$) are 0.304 [0.280,0.329] and for xrsABC-MCMC 0.302 [0.265,0.336]. While the quantiles for pmABC seem very close to the ones from the exchange algorithm, they are actually biased due to the low mixing (see for example \citealp{talts2018validating} for remedies). In fact, a better way to verify the quality of the results is to directly compare the resulting posterior distributions, rather than posterior quantiles. We can do this for all the several runs by separately comparing the forty Wasserstein distances obtained between the corresponding forty posterior from the exchange algorithm and the posteriors from pmABC and, similarly, comparing the distances between the forty posteriors from xrsABC with the ones from the exchange algorithm. The 2d Wasserstein distances where computed using the \texttt{emd2d} function found in the R \texttt{emdist} package. 
Histograms of the distances are in Figure \ref{fig:wasserstein_histograms} and it is clear that xrsABC produces smaller distances, hence these posteriors are closer to the gold-standard exchange algorithm posteriors. Specifically, the median Wasserstein distance for pmABC is 3.37, and for xrsABC is 2.40. 
It could be argued that the above gives reassurance on the quality of ABC inference using stratification when the number of MCMC iterations for each run is limited, but that perhaps pmABC would score more favorably if much longer chains were available, as in this case the poor mixing of the chains could be partially compensated. To address this issue, we ``stack'' the chains produced by the forty runs, to obtain a single long chain for pmABC, and similarly for the exchange algorithm and xrsABC. This way, for each method we have a single posterior resulting from $40\times 2,000=80,000$ draws, which we can easily compare, see Figure \ref{fig:ising_histograms}. We can now see that pmABC produces a more accurate posterior approximation, compared to xrsABC, which is reasonable when a long enough chain is available.

In conclusion, when the number of MCMC iterations is limited, such as when the model simulator is computationally expensive, xrsABC provides a more efficient exploration of the posterior surface  despite using a ``large'' threshold $\delta$, compared to the more slowly moving pmABC. This scenario is further explored in our Lotka-Volterra case study.

\subsection{Lotka-Volterra and sequential Monte Carlo ABC}\label{sec:lv}

Lotka-Volterra is a well studied toy model for testing inference procedures, and has extensively been used in the likelihood-free literature, e.g. \cite{prangle2017adapting}, \cite{papamakarios2016fast}, \cite{bootstrapped-sl}. It describes the time-dynamics of the sizes of two interacting populations $X_1$ and $X_2$, and in its original ecological setting the populations represent predators and prey. However it is also a simple example of biochemical reaction dynamics (Markov jump process) of the kind studied in systems biology. 
Its solution may be simulated exactly using the ``Gillespie algorithm'' \citep{gillespie1977exact},
but it is not possible to evaluate its likelihood. More informations on the model and the summary statistics we used within ABC are in the Supplementary Material.
Data-generating parameters were $\theta_1=1$, $\theta_2=0.008$ and $\theta_3=0.6$.
The simulation starts with initial populations $X_1=50$ and $X_2=100$,
and including the initial values has $n_\mathrm{obs}=32$ measurements for each series,
with the values of $X_1$ and $X_2$ being recorded every 2 time units. 

We compare pmABC-MCMC and rsABC-MCMC to sequential Monte Carlo ABC (ABC-SMC). Briefly, for ABC-SMC we use $N=1,000$ particles, and  following \cite{del2012adaptive} we keep decreasing the threshold $\delta_l$  as long as the acceptance rate is larger than 1.5\%. In our case, at iteration $l=31$ this acceptance rate drops below 1.5\% and hence we consider the draws obtained at the previous iteration $l=30$ as the algorithm output. The value of the threshold producing the reported results is $\delta_{30}=0.859$.
Regarding ABC-MCMC algorithms, we focus on the accuracy of the inference and start ABC-MCMC simulations at the ground-truth parameter values. We propose parameters  via the adaptive MCMC method in \cite{haario2001adaptive}. 
We then run 20,000 iterations of pmABC-MCMC with both $M=1$ and $M=2$ in \eqref{eq:pmABC-lik} (as explained in section \ref{sec:supernova}, in terms of running times, there is no way to perform an exact comparison between rsABC-MCMC and pmABC-MCMC). We considered $\delta=0.6$, since with this value we approximately target a 1\% acceptance rate with $M=2$, whereas using the same $\delta$ found with ABC-SMC would induce a too large acceptance rate.   Results are in Table \ref{tab:lv}. As it is expected in ABC studies, small acceptance rates are targeted to ensure a reasonably precise inference: however, with limited computing time, good exploration of the tails is typically neglected, as discussed for the Ising case study. 
We also report the IAT and ESS. In order to be conservative, we report the largest IAT from the individual chains of the three estimated parameters (the smaller the IAT the better the results). We are also conservative regarding the ESS (the larger the better) for which we report the smallest ESS attained on the three chains. Results show that, as studied in \cite{bornn2017use}, when using pmABC-MCMC it is not really worth to consider $M>1$, see the ESS/min index. 
We also run rsABC-MCMC, but prior to this we performed a study on the effect of different bootstrap schemes, given in Supplementary Material. It turned out that a block bootstrap, with overlapping blocks each having size eight, performed better than other tested alternatives (and, at least for the considered data, much better than the case of non-overlapping blocks).
Notice that, in this experiment, if we were to use a block bootstrap with non-overlapping blocks it would make no sense to resample with $R$ taken larger than 256, since there are at most $4^4=256$ different combinations of four blocks (repetitions being allowed). We use a block bootstrap with overlapping blocks, so we could generate more than 256 different bootstrap datasets, however here we still use $R=256$. Also, unlike in other case studies, here we do not keep the resampled indices constant  throughout iterations, and instead we randomly select them anew for each proposed $\theta$. 
The rsABC-MCMC is not able to run satisfactorily with the same small $\delta$ used for the other algorithms (this is expected, as it will cause some strata to be neglected, this increasing the rejection rate), so we set it to a value eight times larger, that is $\delta_{rs}=8\times 0.6=4.8$. We use the following strata (a justification for this construction is in Supplementary Material):
$\mathcal{D}_1=(0,0.5\delta_{rs})$, $\mathcal{D}_2=[0.5\delta_{rs},\delta_{rs})$,  $\mathcal{D}_3=[\delta_{rs},\infty)$.

Table \ref{tab:lv} shows that inference returned by rsABC-MCMC is  similar to ABC-SMC. By looking at the IAT, rsABC-MCMC requires way fewer iterations to produce an independent sample (about 92 iterations instead of the around 400 of pmABC-MCMC). Clearly, the ESS is also favourable to rsABC-MCMC. Finally, we report the ESS/min, that is the estimated number of independent samples produced per minute. The increase in efficiency is between 5-7-folds in favour of rsABC-MCMC (by comparing the ESS/min). This is an important result, and ideally we could improve on this, if we were to code the resampling part of the algorithm, amounting to very few lines, using some compiled language (e.g. C/C++). We suggest that a better use of rsABC-MCMC is for more computationally expensive case studies, where the model simulation is a bottleneck. We explore this scenario in Supplementary Material, where for a computationally heavier Lotka-Volterra simulation we obtain an efficiency increase between 11-13 times.

\begin{table}
\centering
\resizebox{\textwidth}{!}{  
\begin{tabular}{lccccccccc}
\hline
{} & $\theta_1$ & $\theta_2$ & $\theta_3$ & accept. rate & IAT & ESS & runtime & ESS/min   \\
{} & {} & {} & {} & (\%) & {} & {} & (min) & \\
\hline
true parameters & 1 & 0.008 & 0.6 & & & &\\
ABC-SMC ($\delta=0.859$) & 0.922 [0.719, 1.200] &  0.009 [0.007, 0.012] & 0.682 [0.530, 0.849] & $\approx 1.5\%$ & -- & -- & -- & -- \\
pmABC-MCMC ($\delta=0.6$, $M=1$) & 1.004 [0.853, 1.135]  &  0.008 [0.007, 0.011] &    0.613 [0.526, 0.827] & $\approx  0.5\%$ & 381 & 17.4 & 4.3 & 4.0\\
pmABC-MCMC ($\delta=0.6$, $M=2$) & 0.988 [0.812, 1.188]  &  0.009 [0.007, 0.010]  &  0.636 [0.484, 0.761] & $\approx 1.0\%$ & 403 & 24.6 & 7.6 & 3.2 \\
rsABC-MCMC ($\delta=4.8$) & 1.028 [0.769,  1.362] &  0.009 [0.007,    0.012]    & 0.672 [0.498,   0.909] & $\approx 4.3\%$ & 92 & 117 & 5.7 & \textbf{20.5} 
 \\
\hline
\end{tabular}
}
\caption{\footnotesize{Lotka-Volterra: Mean and 95\% posterior intervals for $\theta$ using several algorithms. Notice rsABC-MCMC uses a much larger $\delta$ than pmABC-MCMC, see main text. Numbers pertaining all ABC-MCMC strategies (including runtime) are based on the last 10,000 samples. The IAT reported is the largest autocorrelation time across chains and ESS is the smallest effective sample size across chains. The best performance is in bold. We do not report the ABC-SMC runtime as it has a completely different algorithmic structure from the MCMC approach and a number of iterations that is not fixed a-priori.}}
\label{tab:lv}
\end{table}

\section{Conclusions}

We have constructed stratified Monte Carlo strategies to substantially reduce the bias induced by data resampling procedures in ABC inference. We have applied our methods to a number of ABC samplers, namely the (generalized) ABC rejection, ABC importance sampling and ABC-MCMC. We have found that, thanks to the combined use of resampling and stratification (i) it is possible to run ABC algorithms using a much larger than usual ABC threshold, this improving the chain mixing (for ABC-MCMC samplers) without sacrificing much in terms of quality of the inference, and (ii) considerably improving the acceptance probability in ABC-rejection and ABC importance sampling, thanks to a reduced variance in the ABC likelihood estimation. However, all these aspects are also connected to the optimal design of the strata, an aspect which is left for future research, though we gave examples on how to decide the size of the strata. 

When the model simulator is fast enough to run, then bootstrapping data (and computing summary statistics) comes with some computational overhead. That is to say, when the code for sampling independently $M>1$ datasets can be easily vectorised (or efficiently parallelised) to allow sampling at a moderate cost, it is reasonable to run  ABC methods that do not use our strategies and instead can exploit the $M$ samples. Otherwise the statistical efficiency might be obfuscated by the increased computational inefficiency \citep{bornn2017use}.
However, in many realistic situations, simulating from the model using $M\gg 1$ is a computational bottleneck and/or producing a vectorised or parallelized code might not be possible. In this case resampling data from a small number of model simulations $M$ is beneficial (when using our strategy, we produced at most $M=2$ samples at each $\theta$). Another aspect that should be taken into account is whether computing the summary statistics comes at a considerable cost. We have assumed the latter not to be the case, and that their computation can be vectorised or parallelised. Clearly, if this is not the case, our approach is still appealing if the model simulator is considerably more expensive to simulate compared to the computation of the  summary statistics for a possibly large number of bootstrapped datasets.
For the Lotka-Volterra case study we have shown that, when considering an expensive model simulator, realistically only a few thousands ABC-MCMC iterations can be performed, this preventing a thorough exploration of the posterior surface. We have shown that our ABC-MCMC using resampling and stratification produces accurate results, and for a computationally intensive Lotka-Volterra case-study it resulted between 11-13 times more efficient than standard pseudo-marginal ABC-MCMC. The astronomical case study in section \ref{sec:supernova} shows that the use of our strategy, with ABC-rejection and ABC importance sampling algorithms, reduces the number of attempted proposals by 2 to 3 times, compared to analogous algorithms not employing our strategy.

\section*{Acknowledgements}
Umberto Picchini acknowledges support by grants from the Swedish Research Council (Vetenskapsr{\aa}det 2019-03924) and the Chalmers AI Research Center (CHAIR). Richard Everitt's work was supported by NERC grants NE/T004010/1 and NE/T00973X/1.

\bibliographystyle{abbrvnat}
\bibliography{biblio}

\begin{thebibliography}{45}
\providecommand{\natexlab}[1]{#1}
\providecommand{\url}[1]{\texttt{#1}}
\expandafter\ifx\csname urlstyle\endcsname\relax
  \providecommand{\doi}[1]{doi: #1}\else
  \providecommand{\doi}{doi: \begingroup \urlstyle{rm}\Url}\fi

\bibitem[Allingham et~al.(2009)Allingham, King, and
  Mengersen]{allingham2009bayesian}
D.~Allingham, R.~King, and K.~Mengersen.
\newblock Bayesian estimation of quantile distributions.
\newblock \emph{Statistics and Computing}, 19\penalty0 (2):\penalty0 189--201,
  2009.

\bibitem[Andrieu and Vihola(2016)]{andrieu2016establishing}
C.~Andrieu and M.~Vihola.
\newblock Establishing some order amongst exact approximations of {MCMC}s.
\newblock \emph{The Annals of Applied Probability}, 26\penalty0 (5):\penalty0
  2661--2696, 2016.

\bibitem[Beaumont et~al.(2002)Beaumont, Zhang, and
  Balding]{beaumont2002approximate}
M.~A. Beaumont, W.~Zhang, and D.~J. Balding.
\newblock Approximate {B}ayesian computation in population genetics.
\newblock \emph{Genetics}, 162\penalty0 (4):\penalty0 2025--2035, 2002.

\bibitem[Bornn et~al.(2017)Bornn, Pillai, Smith, and Woodard]{bornn2017use}
L.~Bornn, N.~S. Pillai, A.~Smith, and D.~Woodard.
\newblock The use of a single pseudo-sample in approximate {B}ayesian
  computation.
\newblock \emph{Statistics and Computing}, 27\penalty0 (3):\penalty0 583--590,
  2017.

\bibitem[Chen and Shao(1999)]{chen1999monte}
M.-H. Chen and Q.-M. Shao.
\newblock {Monte Carlo estimation of Bayesian credible and HPD intervals}.
\newblock \emph{Journal of Computational and Graphical Statistics}, 8\penalty0
  (1):\penalty0 69--92, 1999.

\bibitem[Condon and Matthews(2018)]{condon2018lambdacdm}
J.~Condon and A.~Matthews.
\newblock $\lambda$cdm cosmology for astronomers.
\newblock \emph{Publications of the Astronomical Society of the Pacific},
  130\penalty0 (989):\penalty0 073001, 2018.

\bibitem[Cranmer et~al.(2020)Cranmer, Brehmer, and Louppe]{cranmer2019}
K.~Cranmer, J.~Brehmer, and G.~Louppe.
\newblock The frontier of simulation-based inference.
\newblock \emph{Proceedings of the National Academy of Sciences}, 117\penalty0
  (48):\penalty0 30055--30062, 2020.

\bibitem[Del~Moral et~al.(2012)Del~Moral, Doucet, and Jasra]{del2012adaptive}
P.~Del~Moral, A.~Doucet, and A.~Jasra.
\newblock An adaptive sequential monte carlo method for approximate bayesian
  computation.
\newblock \emph{Statistics and Computing}, 22\penalty0 (5):\penalty0
  1009--1020, 2012.

\bibitem[Drovandi and Pettitt(2011)]{drovandi2011likelihood}
C.~Drovandi and A.~Pettitt.
\newblock Likelihood-free {B}ayesian estimation of multivariate quantile
  distributions.
\newblock \emph{Computational Statistics \& Data Analysis}, 55\penalty0
  (9):\penalty0 2541--2556, 2011.

\bibitem[Everitt(2017)]{bootstrapped-sl}
R.~G. Everitt.
\newblock Bootstrapped synthetic likelihood.
\newblock \emph{arXiv preprint arXiv:1711.05825}, 2017.

\bibitem[Fan and Sisson(2018)]{sissonfan2018}
Y.~Fan and S.~Sisson.
\newblock \emph{Handbook of approximate Bayesian computation}, chapter ABC
  samplers.
\newblock Chapman and Hall/CRC, 2018.
\newblock also available as \texttt{arXiv:1802.09650}.

\bibitem[Fearnhead and Prangle(2012)]{fearnhead-prangle(2011)}
P.~Fearnhead and D.~Prangle.
\newblock Constructing summary statistics for approximate {B}ayesian
  computation: semi-automatic approximate {B}ayesian computation (with
  discussion).
\newblock \emph{Journal of the Royal Statistical Society series B},
  74:\penalty0 419--474, 2012.

\bibitem[Gillespie(1977)]{gillespie1977exact}
D.~T. Gillespie.
\newblock Exact stochastic simulation of coupled chemical reactions.
\newblock \emph{The Journal of Physical Chemistry}, 81\penalty0 (25):\penalty0
  2340--2361, 1977.

\bibitem[Gutmann and Corander(2016)]{gutmann2016bayesian}
M.~U. Gutmann and J.~Corander.
\newblock Bayesian optimization for likelihood-free inference of
  simulator-based statistical models.
\newblock \emph{The Journal of Machine Learning Research}, 17\penalty0
  (1):\penalty0 4256--4302, 2016.

\bibitem[Haario et~al.(2001)Haario, Saksman, Tamminen,
  et~al.]{haario2001adaptive}
H.~Haario, E.~Saksman, J.~Tamminen, et~al.
\newblock An adaptive {M}etropolis algorithm.
\newblock \emph{Bernoulli}, 7\penalty0 (2):\penalty0 223--242, 2001.

\bibitem[H{\"a}rdle et~al.(2003)H{\"a}rdle, Horowitz, and
  Kreiss]{hardle2003bootstrap}
W.~H{\"a}rdle, J.~Horowitz, and J.-P. Kreiss.
\newblock Bootstrap methods for time series.
\newblock \emph{International Statistical Review}, 71\penalty0 (2):\penalty0
  435--459, 2003.

\bibitem[Jennings and Madigan(2017)]{jennings2017astroabc}
E.~Jennings and M.~Madigan.
\newblock astroabc: an approximate bayesian computation sequential monte carlo
  sampler for cosmological parameter estimation.
\newblock \emph{Astronomy and computing}, 19:\penalty0 16--22, 2017.

\bibitem[Karabatsos and Leisen(2018)]{karabatsos2017approximate}
G.~Karabatsos and F.~Leisen.
\newblock An approximate likelihood perspective on {ABC} methods.
\newblock \emph{Statistics Surveys}, 12:\penalty0 66--104, 2018.

\bibitem[Kreiss and Paparoditis(2011)]{kreiss2011bootstrap}
J.-P. Kreiss and E.~Paparoditis.
\newblock Bootstrap methods for dependent data: A review.
\newblock \emph{Journal of the Korean Statistical Society}, 40\penalty0
  (4):\penalty0 357--378, 2011.

\bibitem[Kunsch(1989)]{kunsch1989jackknife}
H.~R. Kunsch.
\newblock The jackknife and the bootstrap for general stationary observations.
\newblock \emph{The annals of Statistics}, pages 1217--1241, 1989.

\bibitem[Lahiri(2013)]{lahiri2013resampling}
S.~N. Lahiri.
\newblock \emph{Resampling methods for dependent data}.
\newblock Springer Science \& Business Media, 2013.

\bibitem[Lenormand et~al.(2013)Lenormand, Jabot, and
  Deffuant]{lenormand2013adaptive}
M.~Lenormand, F.~Jabot, and G.~Deffuant.
\newblock Adaptive approximate {B}ayesian computation for complex models.
\newblock \emph{Computational Statistics}, 28\penalty0 (6):\penalty0
  2777--2796, 2013.

\bibitem[Lintusaari et~al.(2017)Lintusaari, Gutmann, Dutta, Kaski, and
  Corander]{lintusaari2017fundamentals}
J.~Lintusaari, M.~U. Gutmann, R.~Dutta, S.~Kaski, and J.~Corander.
\newblock Fundamentals and recent developments in approximate {B}ayesian
  computation.
\newblock \emph{Systematic biology}, 66\penalty0 (1):\penalty0 e66--e82, 2017.

\bibitem[Marjoram et~al.(2003)Marjoram, Molitor, Plagnol, and
  Tavar{\'e}]{marjoram2003markov}
P.~Marjoram, J.~Molitor, V.~Plagnol, and S.~Tavar{\'e}.
\newblock Markov chain {M}onte {C}arlo without likelihoods.
\newblock \emph{Proceedings of the National Academy of Sciences}, 100\penalty0
  (26):\penalty0 15324--15328, 2003.

\bibitem[Murray et~al.(2006)Murray, Ghahramani, and MacKay]{Murray2006}
I.~Murray, Z.~Ghahramani, and D.~J.~C. MacKay.
\newblock {MCMC for doubly-intractable distributions}.
\newblock In \emph{Proceedings of the Twenty-Second Conference on Uncertainty
  in Artificial Intelligence (UAI2006)}, pages 359--366, 2006.

\bibitem[Nott et~al.(2019)Nott, Drovandi, and Kohn]{nott2019bayesian}
D.~J. Nott, C.~Drovandi, and R.~Kohn.
\newblock Bayesian inference using synthetic likelihood: asymptotics and
  adjustments.
\newblock \emph{arXiv preprint arXiv:1902.04827}, 2019.

\bibitem[Owen(2013)]{owen}
A.~B. Owen.
\newblock \emph{Monte Carlo theory, methods and examples}.
\newblock 2013.
\newblock \url{http://statweb.stanford.edu/~owen/mc/}, retrieved 15 April 2018.

\bibitem[Papamakarios and Murray(2016)]{papamakarios2016fast}
G.~Papamakarios and I.~Murray.
\newblock Fast $\varepsilon$-free inference of simulation models with bayesian
  conditional density estimation.
\newblock In \emph{Advances in Neural Information Processing Systems}, pages
  1028--1036, 2016.

\bibitem[Picchini(2018)]{picchini2018likelihood}
U.~Picchini.
\newblock Likelihood-free stochastic approximation {EM} for inference in
  complex models.
\newblock \emph{Communications in Statistics-Simulation and Computation}, 2018.
\newblock doi:10.1080/03610918.2017.1401082.

\bibitem[Picchini and Forman(2016)]{picchini2016accelerating}
U.~Picchini and J.~L. Forman.
\newblock Accelerating inference for diffusions observed with measurement error
  and large sample sizes using approximate {B}ayesian computation.
\newblock \emph{Journal of Statistical Computation and Simulation}, 86\penalty0
  (1):\penalty0 195--213, 2016.

\bibitem[Picchini et~al.(2020)Picchini, Simola, and
  Corander]{picchini2020adaptive}
U.~Picchini, U.~Simola, and J.~Corander.
\newblock Sequentially guided {MCMC} proposals for synthetic likelihoods and
  correlated synthetic likelihoods.
\newblock \emph{arXiv preprint arXiv:2004.04558}, 2020.

\bibitem[Plummer et~al.(2006)Plummer, Best, Cowles, and Vines]{coda}
M.~Plummer, N.~Best, K.~Cowles, and K.~Vines.
\newblock {CODA}: Convergence diagnosis and output analysis for {MCMC}.
\newblock \emph{R News}, 6\penalty0 (1):\penalty0 7--11, 2006.
\newblock URL \url{https://journal.r-project.org/archive/}.

\bibitem[Politis and Romano(1994)]{politis1994stationary}
D.~N. Politis and J.~P. Romano.
\newblock The stationary bootstrap.
\newblock \emph{Journal of the American Statistical association}, 89\penalty0
  (428):\penalty0 1303--1313, 1994.

\bibitem[Prangle(2017)]{gk}
D.~Prangle.
\newblock gk: An {R} package for the g-and-k and generalised g-and-h
  distributions.
\newblock \emph{arXiv:1706.06889}, 2017.

\bibitem[Prangle et~al.(2017)]{prangle2017adapting}
D.~Prangle et~al.
\newblock Adapting the {ABC} distance function.
\newblock \emph{Bayesian Analysis}, 12\penalty0 (1):\penalty0 289--309, 2017.

\bibitem[Price et~al.(2018)Price, Drovandi, Lee, and Nott]{price2018bayesian}
L.~F. Price, C.~C. Drovandi, A.~Lee, and D.~J. Nott.
\newblock Bayesian synthetic likelihood.
\newblock \emph{Journal of Computational and Graphical Statistics}, 27\penalty0
  (1):\penalty0 1--11, 2018.

\bibitem[Quilis(2021)]{quilis}
E.~M. Quilis.
\newblock Bootstrapping time series.
\newblock MATLAB Central File Exchange, retrieved January 12, 2021, 2021.
\newblock
  \url{https://www.mathworks.com/matlabcentral/fileexchange/53701-bootstrapping-time-series}.

\bibitem[Rayner and MacGillivray(2002)]{rayner2002numerical}
G.~D. Rayner and H.~L. MacGillivray.
\newblock Numerical maximum likelihood estimation for the g-and-k and
  generalized g-and-h distributions.
\newblock \emph{Statistics and Computing}, 12\penalty0 (1):\penalty0 57--75,
  2002.

\bibitem[Rubinstein and Kroese(2016)]{rubinstein2016simulation}
R.~Y. Rubinstein and D.~P. Kroese.
\newblock \emph{Simulation and the {Monte Carlo} method}.
\newblock John Wiley \& Sons, r edition, 2016.

\bibitem[Simola et~al.(2021)Simola, Cisewski-Kehe, Gutmann, and
  Corander]{simola2021adaptive}
U.~Simola, J.~Cisewski-Kehe, M.~U. Gutmann, and J.~Corander.
\newblock Adaptive approximate {B}ayesian computation tolerance selection.
\newblock \emph{Bayesian Analysis}, 16\penalty0 (2):\penalty0 397--423, 2021.

\bibitem[{Statisticat} and {LLC.}(2018)]{laplaces}
{Statisticat} and {LLC.}
\newblock \emph{LaplacesDemon Examples}, 2018.
\newblock URL
  \url{https://web.archive.org/web/20150206004624/http://www.bayesian-inference.com/software}.
\newblock R package version 16.1.1.

\bibitem[Talts et~al.(2018)Talts, Betancourt, Simpson, Vehtari, and
  Gelman]{talts2018validating}
S.~Talts, M.~Betancourt, D.~Simpson, A.~Vehtari, and A.~Gelman.
\newblock Validating bayesian inference algorithms with simulation-based
  calibration.
\newblock \emph{arXiv preprint arXiv:1804.06788}, 2018.

\bibitem[Wilkinson(2013)]{wilkinson-abc-summaries}
D.~J. Wilkinson.
\newblock Summary stats for {ABC}, 2013.
\newblock
  \url{https://darrenjw.wordpress.com/2013/09/01/summary-stats-for-abc/},
  retrieved 23 March 2019.

\bibitem[Wiqvist et~al.(2019)Wiqvist, Mattei, Picchini, and
  Frellsen]{wiqvist2019partially}
S.~Wiqvist, P.-A. Mattei, U.~Picchini, and J.~Frellsen.
\newblock Partially exchangeable networks and architectures for learning
  summary statistics in approximate bayesian computation.
\newblock In \emph{International Conference on Machine Learning}, pages
  6798--6807. PMLR, 2019.

\bibitem[Wood(2010)]{wood2010statistical}
S.~N. Wood.
\newblock Statistical inference for noisy nonlinear ecological dynamic systems.
\newblock \emph{Nature}, 466\penalty0 (7310):\penalty0 1102, 2010.

\end{thebibliography}

\clearpage
\newpage

\begin{center}
 \LARGE{Supplementary Material}   
\end{center}

\section{Unbiasedness of $\hat{\mu}_\mathrm{strat}$}

The estimator $\hat{\mu}_\mathrm{strat}$ in (5) in the main text is easily shown to be unbiased, see chapter 8 in \cite{owen}. This fact is proved below for ease of access. Recall, when using (5) we assume the ability to simulate directly from each given stratum, therefore we denote with $x\sim p_j$ a simulation from stratum $\mathcal{D}_j$, where $p_j(x)=p(x|x\in\mathcal{D}_j)=\omega_j^{-1}p(x)\mathbb{I}_{x\in \mathcal{D}_j}$. Then we have

\begin{align*}
\mathbb{E}(\hat{\mu}_{\mathrm{strat}})&= \sum_{j=1}^J\omega_j\mathbb{E}\biggl(\frac{1}{\tilde{n}_j}\sum_{i=1}^{\tilde{n}_j}f(x_{ij})\biggr)=\sum_{j=1}^J\omega_j\int_{\mathcal{D}_j}f(x)p_j(x)dx\\
&=\sum_{j=1}^J\int_{\mathcal{D}_j}f(x)p(x)dx
=\int_{\mathcal{D}}f(x)p(x)dx=\mu.
\end{align*}

\section*{Efficiency of the averaged likelihood approach}

\begin{figure}[ht]
    \centering
    \includegraphics[width=9cm,height=5cm]{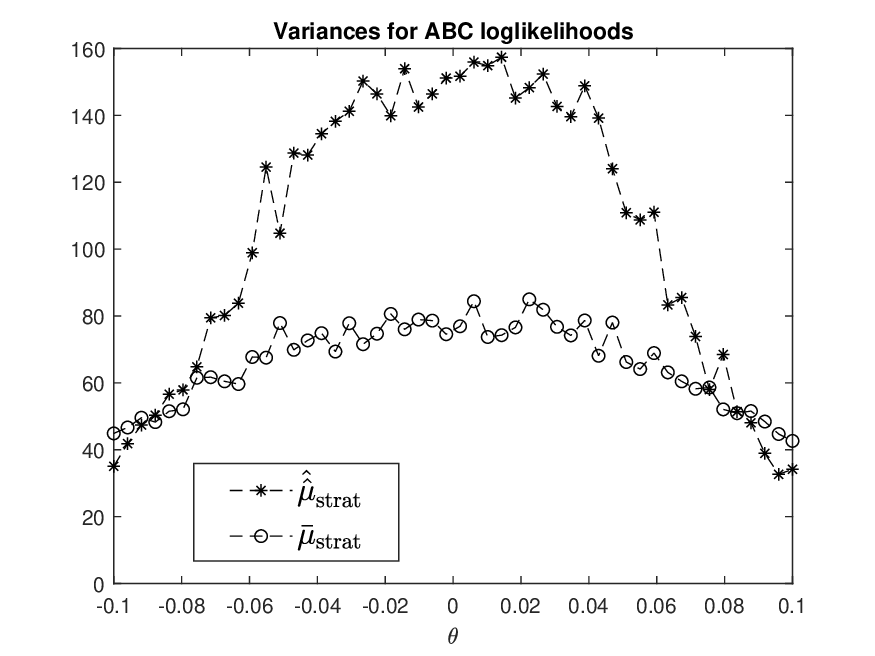}
    \caption{\footnotesize{1D Gauss example: variance of $\hat{\hat{\mu}}_\mathrm{strat}$ (*) and variance of $\bar{\mu}_\mathrm{strat}$ (circles).}}
    \label{fig:gauss-variances-avglik-nonavglik}
\end{figure}

In section 4.2 of the main paper we considered averaging two likelihoods obtained via stratification. Here we show that the approach is promising in terms of variance reduction. Here we consider the same data studied in the simple Gaussian case study. We compute both $\hat{\hat{\mu}}_\mathrm{strat}$ (which is the same as $\hat{\hat{\mu}}_\mathrm{strat}^{(1)}$, see section 4.2) and $\bar{\mu}_\mathrm{strat}$ at fifty equispaced values taken in the interval $\theta\in [-0.1,0.1]$.  For each of the fifty considered values of $\theta$, both approximate likelihoods are computed for 1,000 independent repetitions, using $\delta=3\times 10^{-4}$, $R=500$ and the same three strata as considered in the main text. Figure \ref{fig:gauss-variances-avglik-nonavglik} gives the variances of the corresponding likelihood approximations obtained over the 1,000 repetitions (recall the ground-truth parameter for the data generating process is $\theta=0$). 
We deduce that in the central values of the space, around 50\% decrease in the variance is obtained when using $\bar{\mu}_\mathrm{strat}$ compared to using $\hat{\hat{\mu}}_\mathrm{strat}$. 

\section{Computational considerations for rABC-MCMC and rsABC-MCMC}

Notice, for rABC-MCMC and with notation specific for a vector of iid data (generalizations are possible depending on the nature of the data), whenever we wish to generate say $R$ resampled datasets, we create a matrix of positive integers $u\equiv [u_{r,i}]_{r=1,...,R;i=1,...,n_\mathrm{obs}}$, where each $u_{r,i}$ is obtained by uniform random sampling with replacement from $\{1,...,n_\mathrm{obs}\}$. Therefore $u$ collects the indeces of the values of $x^*$ that have been resampled. Then, in order to obtain a computational saving, we reuse the same matrix $u$ across the rABC-MCMC iterations, that is the $u$ are never modified. This strategy can also be implemented for other (non MCMC) ABC samplers.
For rsABC-MCMC we do the same (again, when data are iid, for ease of illustration), except that we produce two distinct matrices, $u_1$ and $u_2$, one to be used for summaries employed in the determination of the $n_j$ and the other for the summaries used when estimating the $\omega_j$. We keep these two matrices constant during rsABC-MCMC.

\section{Supernova example}

\subsection{Data generation and summary statistics}\label{secsupp:supernova-datagen}

Here we describe how to simulate a generic dataset. We generate $10^4$ variates $u_1,...,u_{10^4}$, independently sampled from a truncated Gaussian $u_j\sim \mathcal{N}_{[0.01,1.2]}(0.5,0.05^2)$ ($j=1,...,10^4$), where $\mathcal{N}_{[a,b]}(m,\sigma^2)$ denotes a Gaussian distribution with mean $m$ and variance $\sigma^2$, truncated to the interval $[a,b]$. The $u_j$ are then binned into 20 intervals of equal width (essentially the bins of an histogram constructed on the $u_j$), then the 20 centres of the bins are obtained and these centres are the ``redshifts'' $z_1,...,z_{20}$. Then for each $z_i$ we compute the distance modulus $\mu_i$ via (15) in the main text. Therefore, each simulation from the model requires first the generation of the 10,000 truncated Gaussians, then their binning and the calculations of the twenty $\mu_i$. 
We consider $s=(\mu_1,...,\mu_{20})$ as summary statistics, which are in ascending order (i.e. $\mu_j\leq \mu_{j+1}$ for all $j$): this is because of how the generative model is formulated, to return ``bins centres'' from an histogram, which is of course a sequence of increasing numbers. 
Notice, when data are simulated as illustrated above, $s$ is a trivial summary statistic, in that $(\mu_1,...,\mu_{20})$ is the data itself (since both the $u_j$ and the $z_i$ do not depend on $\theta$). It is therefore important to note that, when using our stratified Monte Carlo procedure in this specific example, once the simulated  $(\mu_1,...,\mu_{20})$ have been resampled, \textbf{we must sort each resampled vector in ascending order, before computing the ABC distances} (since ``observed summaries'' are sorted, as previously noted).

\subsection{Supernova example: generalized ABC-rejection with stratified Monte Carlo}\label{secsupp:supernova-strata}

\begin{figure}
    \centering
    \includegraphics[scale=0.5]{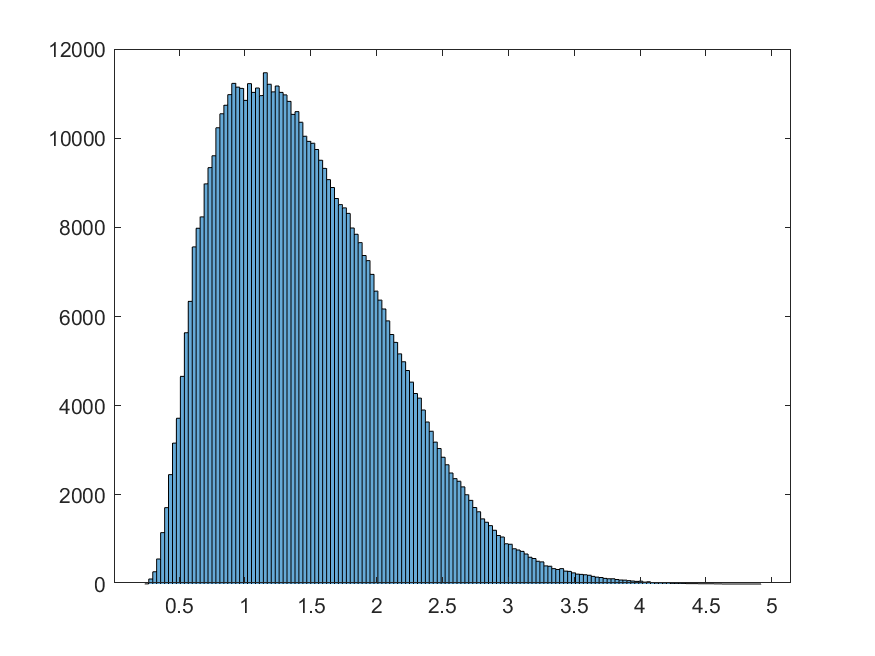}
    \caption{Supernova: histogram of distances simulated from the prior predictive distribution incorporating data bootstrapping (subsampled for ease of visualization).}
    \label{fig:astro-dist_rejection}
\end{figure}

For the supernova example we run both algorithm 1 and algorithm 2 from the main text, using $M=1$ or $M=2$ for the former one, and $R=3,000$ in the second one. 
For this example, when using the generalized ABC-rejection algorithm we used strata $\mathcal{D}_1=(0,0.45)$, $\mathcal{D}_2=(0.45,0.55)$, $\mathcal{D}_3=(0.55,\infty)$. Their choice was guided by distances produced from 2,000 simulations from the prior predictive distribution using bootstrapped simulated data, see Figure \ref{fig:astro-dist_rejection}. More in detail, we simulated 2,000 parameters $\theta^*\sim \pi(\theta)$, and at each parameter we produced a corresponding model simulation $x^*\sim p(x|\theta^*)$ (also corresponding, for this model, to a simulated $s^*$ after sorting is applied, see section \ref{secsupp:supernova-datagen}), which we resample with replacement to produce 3,000 bootstrapped datasets at each $\theta^*$. 
Overall we have a total of $3,000\times 2,000=6\cdot 10^6$ distances, reported in Figure \ref{fig:astro-dist_rejection}. Each distance is computed as $d^*=(\sum_{i=1}^{20} (s^{i*}-s^i)^2)^{1/2}$, since each $s$ has dimension twenty (notice here we take $\Sigma$ to be the identity matrix since here summaries have the same magnitude). Of course, since the distances are from the prior predictive, only a very small fraction of those corresponds to parameters close to the one that generated data. Since ABC-rejection algorithms sample from the prior of the parameters, which is typically very different from the posterior, we build strata using not-too-small quantiles from the produced distances (or we would frequently incur into empty strata). In this case, the first quantile of the distances distribution is approximately 0.45, and the third quantile is 0.55, which we used to define our strata (in the hypothetical scenario of having better information on the likely values of ground truth parameters, we would of course have more informative priors and hence smaller generated distances that would allow us to take even smaller quantiles).

Now that strata have been obtained, we can run a short pilot that is exclusively meant to collect values of the ABC log-likelihood $\log\hat{\hat{\mu}}_\mathrm{strat}$ evaluated at several proposed values of $\theta$. From this collection, we then take $c=\max\{\exp(\log\hat{\hat{\mu}}_\mathrm{strat})\}$ (or $c_2=\max\{\exp(\log{\bar{\mu}}_\mathrm{strat})\}$), that is the largest value among all obtained likelihoods (or possibly a slightly larger one for good measure), and plug the latter into algorithm 2.

At a first glance, it may appear as implementing stratified Monte Carlo within ABC-rejection requires additional preparatory steps (i.e. to define appropriate strata or compute $c$). However, realistically, it has to be considered that we typically need to produce some pilot ABC run, at the very least to assess the variation of each of the $n_s$ elements that compose the vector of summaries and that allow an appropriate normalization of both the observed summaries and the simulated ones (this essentially amounts to find the diagonal entries in matrix $\Sigma$, see section 4 in the main text). Summaries simulated in such preparatory run could be recycled (after normalization) to compute quantiles of the distances, and hence obtaining information to produce the strata would be a simple by-product of preparatory (and typically necessary) runs.  Pilot runs would also be necessary for other reasons than the ones we touched upon, say to construct summary statistics ``semi-automatically'', as in \cite{fearnhead-prangle(2011)} and \cite{wiqvist2019partially}.  Generally ABC algorithms are hardly really ``plug-and-play'', even though they are often presented as such.

For the case of the supernova example we do not need any summaries normalization, since the summaries vary on the same scale, but that's uncommon.

\subsection{ABC importance sampling algorithms}

For the supernova example, we also implemented ABC importance sampling. In particular, we refer to the one given in \cite{sissonfan2018}, where an importance sampling density $g(\theta)$ is assumed available to the user as input to the algorithm. The version in Table \ref{alg:abc-is} merges algorithms denoted as Algorithm 2 and 3 in \cite{sissonfan2018} and produces $N$ weighted posterior samples $(\theta^{(i)},\tilde{w}^{(i)})$, $i=1,...,N$.
\begin{table}[tbh]
\small
\caption{\bf ABC importance sampling (ABC-IS)}
\label{alg:abc-is}
\noindent For $i=1, \ldots, N$:
\begin{enumerate}
\item Generate $\theta^{(i)}\sim g(\theta)$ from importance sampling density $g$.
\item Generate $M$ datasets $x^{(i,1)},...,x^{(i,M)}\sim p(x|\theta^{(i)})$ iid from the model. 
\item Compute summary statistic $s^{(i,m)}=S(x^{(i,m)})$, $m=1,...,M$.
\item Sample $u\sim U(0,1)$, and if $u<\sum_{m=1}^M K_\delta(s^{(i,m)},s)/(M\cdot K_{\delta,0})$ set $\tilde{w}^{(i)}=\pi(\theta^{(i)})/g(\theta^{(i)})$, else go to step 1.
\end{enumerate}

\noindent {\it Output:}\\
A set of weighted parameter vectors $(\theta^{(1)},\tilde{w}^{(1)}),\ldots,(\theta^{(N)},\tilde{w}^{(N)})$ $\sim$ $\pi_{\delta}(\theta|s)$.
\end{table}

If we consider $K_\delta(s^{(i)},s)\equiv K_\delta(||s^{(i)}-s||)$ (where $||\cdot||$ could be e.g. the Euclidean or Mahalanobis distance), then we define $K_{\delta,0}\equiv K_\delta(0)$.
We can then compute the ABC posterior mean as $E(\theta|s)\approx \sum_{i=1}^N W^{(i)}\theta^{(i)}$ (with $W^{(i)}=\tilde{w}^{(i)}/\sum_{j=1}^N \tilde{w}^{(j)}$) and posterior quantiles by using eq. (3.6) in \cite{chen1999monte}. Of course, we need an importance sampler $g(\theta)$ first. It is not central for our work to determine how to construct such $g(\theta)$, as this is application specific. For the supernova example, we take the draws returned by the generalized ABC-rejection algorithm, we compute their sample mean $m$ and sample covariance $C$, and define $g(\theta)=\mathcal{N}(m,2 C)$, that is we define a multivariate Gaussian importance sampler. Also, we use $M=1$ and $M=2$ and $\delta=0.15$ and obtain $N=1,000$ posterior samples.

A version of ABC-IS incorporating stratified Monte Carlo is easily constructed, and denoted sABC-IS, see Table \ref{alg:sabc-is}. 
\begin{table}
\small
\caption{\bf Stratified Monte Carlo ABC importance sampling (sABC-IS)}
\label{alg:sabc-is}
\noindent For $i=1, \ldots, N$:
\begin{enumerate}
\item Generate $\theta^{(i)}\sim g(\theta)$ from importance sampling density $g$.
\item 
     generate \textit{once} $x^{(i)} \sim p(x| \theta^{(i)})$;
     \item Produce $R$ bootstrapped datasets from $x^{(i)}$. Call these $\{x^{r*}\}_{r=1:R}$; compute corresponding summaries and  obtain the $n_j$, $j=1,...,J$;
     \item Go back to 1 as soon as some $n_j=0$ otherwise continue;
     \item Generate \textit{once} 
     $x^{(i)'} \sim p(x| \theta^{(i)})$.
     \item  Produce $R$ bootstrapped datasets from $x^{(i)'}$. Call these $\{x^{r'}\}_{r=1:R}$; compute corresponding summaries and  estimate the $\omega_j$, $j=1,...,J$;
     \item compute $\hat{\hat{\mu}}_{\mathrm{strat}}$ (or compute ${\bar{\mu}}_{\mathrm{strat}}$; in the step below we write $\hat{\hat{\mu}}_{\mathrm{strat}}$ as an example );
\item Sample $u\sim U(0,1)$, and if $u<\hat{\hat{\mu}}_{\mathrm{strat}}/\hat{\hat{\mu}}_{\mathrm{strat},0}$ set $\tilde{w}^{(i)}=\pi(\theta^{(i)})/g(\theta^{(i)})$, else go to step 1.
\end{enumerate}

\noindent {\it Output:}\\
A set of weighted parameter vectors $(\theta^{(1)},\tilde{w}^{(1)}),\ldots,(\theta^{(N)},\tilde{w}^{(N)})$ $\sim$ $\pi_{\delta}(\theta|s)$.
\end{table}
Notice, whenever we reach step 8 in sABC-IS surely all $n_j$ are strictly positive for that iteration, hence
\begin{equation*}
\hat{\hat{\mu}}_{\mathrm{strat},0}=\sum_{j=1}^J \biggl\{\frac{\hat{\omega}_j}{n_j} \sum_{i=1}^{n_j} K_\delta(0)\biggr\},
\end{equation*}
 and for the specific case of Gaussian kernels we have that 
$\hat{\hat{\mu}}_{\mathrm{strat},0}= \delta^{-n_s}$.

For the supernova example, we do not reuse the same strata employed for the ABC rejection algorithm. Since here we did make use of the results from the ABC rejection to construct an importance sampler, we simulate distances from summaries produced via parameters drawn from the importance sampler $g(\cdot)$, and compute the quantiles at probability levels 0.03\% and 1\% from these new distances, resulting in $\mathcal{D}_1=(0,0.29)$, $\mathcal{D}_2=(0.29,0.42)$, $\mathcal{D}_3=(0.42,\infty)$.
Then, we ran sABC-IS with $R=3,000$ (to obtain $\hat{\hat{\mu}}_{\mathrm{strat}}$),  $N=1,000$ and enlarged the threshold to five times the value used for ABC-IS, that is here we used $\delta=0.75$.

\section{Lotka-Volterra study}

\subsection{The model setup}

The Lotka-Volterra model describes how the
number of individuals in two populations (one of predators, the other
of prey) change over time. 
Here $X_1$ represent the number of predators and $X_2$ the number
of prey. The following three reactions may take place (this is just a possible example, \citealp{papamakarios2016fast} consider four reactions instead): (i) a prey may be born, with rate $\theta_{1}X_2$, increasing $X_2$ by one; (ii) the predator-prey interaction in which $X_1$ increases by one and $X_2$
decreases by one, with rate $\theta_{2}X_1X_2$; (iii) a predator may die, with rate $\theta_{3}X_1$, decreasing $X_1$ by one. For
ABC inference, we followed the summary
statistics used in \cite{wilkinson-abc-summaries} and \cite{papamakarios2016fast}, that is a 9-dimensional vector composed of the sample mean, natural logarithm of the sample variance
and first two autocorrelations (lag 1 and lag 2) of each of the two time series, together with
the Pearson correlation between them. Priors were set uniform $\log\theta_j\sim U(-5,2)$ on the log-transformed parameters, $j=1,2,3$.

\subsection{ABC Sequential Monte Carlo used in the Lotka-Volterra study}

For ABC-SMC  we use the version in \cite{sissonfan2018} that extends the ABC-SMC found in \cite{del2012adaptive} to a generic kernel (\citealp{del2012adaptive} only consider the identity kernel).  This way we can use ABC-SMC with a Gaussian kernel as we did in other applications. This generalization is reported in Table \ref{tab:abc-smc} for ease of access.
Before executing any of the discussed algorithms, we produced 5,000 simulations from the prior predictive distribution of the model. This means that we simulated 5,000 datasets $\{x^i=(x_1^i,x_2^i)\}_{i=1}^{5000}$ with $x^i\sim p(x)=\int p(x|\theta)\pi(\theta)d\theta$, and for each $x^i$ we computed corresponding summaries $s^i=s(x^i)$. We used these summaries in two ways: (i) to obtain a scaling matrix $\Sigma$ for the ABC distances (see section 4 in the main paper) and (ii) to obtain an histogram of ABC distances, which is useful to set the strata for rsABC-MCMC (see section \ref{sec:setting-lv-strata} in this document).

In Table \ref{tab:abc-smc} we report the ABC sequential Monte Carlo (ABC-SMC) algorithm, as described in \cite{sissonfan2018}, except that here we have removed a typo appearing in the original version, where the weights $w_l^{(i)}$ in step 1 were made depend on the ratio of prior densities. Additionally, \cite{sissonfan2018} consider a pseudo-marginal version where, conditionally on each parameter (``particle''), several independent datasets are produced ($T$ datasets in their notation, $M$ datasets in our notation). Given that ABC-SMC can be computer intensive, in our experiments we always generate a single dataset ($M=1$) conditionally on a given particle, hence for simplicity of reading we remove the dependency on $M$. 

In our experiments we run ABC-SMC using the prior as initial sampler, i.e. $q(\theta_0^{(i)})\equiv  \pi(\theta_0^{(i)})$. Prior to starting the inference we compute distances from prior predictive simulations and take the median of those as initial $\delta$. Then we set $\gamma=0.8$, which implies a relatively slow decrease of $\delta_l$ ($\gamma\in(0,1)$ and the closer $\gamma$ is to 1 the slower the decrease) and resample when the effective sample size $ESS(w^{(1)}_l,\ldots,w^{(N)}_l)$ is smaller than $E=N/2$. Notice $ESS(w^{(1)}_l,\ldots,w^{(N)}_l)=\bigl(\sum_{i=1}^N (W_l^{(i)})^2\bigr)^{-1}$ where the $W_l^{(i)}$ are normalized weights. The ABC threshold is updated at each iteration $l$ as described in step 1, which essentially means solving for an unknown $\delta_l$ the equation $$h(\delta_l)=ESS(w^{(1)}_l,\ldots,w^{(N)}_l;\delta_l)-\gamma \cdot ESS(w^{(1)}_{l-1},\ldots,w^{(N)}_{l-1};\delta_{l-1}),$$ i.e. select $\delta_{l}$ so that $h(\delta_{l})=0$. This can be accomplished using some numerical approach for finding roots of nonlinear functions. We used MATLAB's \texttt{fzero} to search for $\delta_l\in[0.5\cdot\delta_{l-1},\delta_{l-1}]$. Note that the ESS in this section has nothing to do with the ESS reported e.g. in Table 2 in the main text. There, the ESS is a measure of the efficiency of an MCMC sampler in producing nearly-independent samples, as computed via the R \texttt{coda} package \citep{coda}.

\begin{table}[tbh]
\scriptsize
\caption{ABC-SMC}
\noindent {\it Inputs:}
\begin{itemize}
\item A kernel function $K_\delta(u)$, number of particles $N>0$.
\item An initial sampling density $q(\theta)$ and sequence of proposal densities $q_l(\theta'|\theta)$, $l=1,\ldots,L$.
\item A value $\gamma\in(0,1)$ to control the effective sample size.
\item A low dimensional vector of summary statistics $s=S(y)$.
\end{itemize}

\noindent {\it Initialise:}\\
For $i=1,\ldots,N$: 
\begin{itemize}
\item Generate $\theta^{(i)}_0 \sim q(\theta)$ from the initial sampling distribution.
\item Generate $y^{(i)}_0 \sim p(y|\theta^{(i)}_0)$ and compute summary statistics $s^{(i)}_0=S(y^{(i)}_0)$.
\item Compute weights 
$w^{(i)}_0= 
\pi(\theta_0^i)/q(\theta^{(i)}_0)$, and set $l=1$.
\end{itemize}

\noindent {\it Sampling:}
\begin{enumerate}
\item \label{alg8:step1} Reweight: Determine $\delta_l$ via optimization, such that $\mathrm{ESS}(w^{(1)}_l,\ldots,w^{(N)}_l)=\gamma \cdot \mathrm{ESS}(w^{(1)}_{l-1},\ldots,w^{(N)}_{l-1})$
where  
\[	w^{(i)}_l=w^{(i)}_{l-1}\frac{K_{\delta_l}(s^{(i)}_{l-1},s)}
	{K_{\delta_{l-1}}(s^{(i)}_{l-1},s)}.
\]
One the optimal $\delta_l$ is found, obtain the corresponding values of $w_l^{(i)}$ based on the previous formula.
Set $\theta^{(i)}_l=\theta^{(i)}_{l-1}$ and $s^{(i)}_l=s^{(i)}_{l-1}$  for $i=1,\ldots,N$. Go to step 2.

\item \label{alg8:step2} Resample: If $\mathrm{ESS}(w^{(1)}_l,\ldots,w^{(N)}_l)<E$ (otherwise go immediately to step 3)  resample $N$ particles $\{\theta^{(i)}_l,s^{(i)}_l\}$ having normalised weights $W_l^{(i)}$, i.e. $W_l^{(i)}=w_l^{(i)}/\sum_{j=1}^Nw_l^{(j)}$. Then reset $w_l^{(i)}=1/N$, recompute and store $\mathrm{ESS}(w_l^{(1)},\ldots,w_l^{(N)})$ and go to step 3.

\item \label{alg8:step 3} Move: For $i=1,\ldots,N$: \\
\hspace{0.6cm} If $w_l^{(i)}>0$ do
\begin{itemize}
\item Generate $\theta'\sim q_l(\theta|\theta^{(i)}_l)$, $y'\sim p(y|\theta')$ 
and compute $s'=S(y')$.

\item Accept $\theta'$ with probability
\begin{equation}
	\min\left\{1,\frac{K_{\delta_l}(s',s)\pi(\theta')q(\theta^{(i)}_l|\theta')}
	{K_{\delta_{l}}(s^{(i)}_{l},s)\pi(\theta^{(i)}_{l})q(\theta'|\theta^{(i)}_l)}\right\}\label{eq:mh-ratio}
\end{equation}
and set $\theta_l^{(i)}=\theta'$, $s^{(i)}_l=s'$.
\end{itemize}
end.
\item Increment $l=l+1$. If stopping rule is not satisfied, go to step \ref{alg8:step1}.
\end{enumerate}

\noindent {\it Output:}\\
A set of weighted parameter vectors $(\theta^{(1)}_L,w^{(1)}_L),\ldots,(\theta^{(N)}_L,w^{(N)}_L)$ drawn from $\pi_{ABC}(\theta|s)\propto \pi(\theta) \int K_{\delta_L}(s',s)p(s'|\theta)ds'$.
\label{tab:abc-smc}
\end{table}

\subsection{Lotka-Volterra: comparing bootstrap strategies for time series}

We compare the (i) block bootstrap with non-overlapping blocks (NOBB) (\citealp{kunsch1989jackknife}), the (ii)  block bootstrap with overlapping blocks (OBB), and (iii) the stationary bootstrap (SB)( \citealp{politis1994stationary}), for the dataset we considered in the Lotka-Volterra study. Additionally, since in the other case-studies we assumed that the indeces for the resampled units were kept constant across the MCMC iterations, we also consider the effect on NOBB of keeping these indeces fixed or randomize them. For NOBB with fixed indeces we used our own implementation, and for the other cases we used \cite{quilis} for MATLAB.
We always resample a dataset $R=256$ times, and this is an upper limit only for the case named NOBB-fixed below, but we use the same bootstrap size across all methods for consistency of comparison.

The setup for the comparison is as follows: for each bootstrap method, we simulate independently 1,000 pairs of parameters and datasets from the prior predictive distribution. That is we draw a $\theta^*$ from the prior, and conditionally on $\theta^*$ we simulate a dataset $x^*$, and we repeat this independently for 1,000 times. For each simulated $x^*$ we bootstrap 256 datasets, and obtain corresponding 256 summaries $s^*$ and hence corresponding distances $d^*$ (see section \ref{secsupp:supernova-strata}). Therefore we have in total $256\times 1,000$ distances from the prior predictive. From the distribution of these distances we store, for each bootstrap strategy, five percentiles, namely: the 0.1th, 0.2th,...,0.5th percentile of the 250,000 distances (we focus on small percentiles because the distances are obtained from the prior predictive, hence the vast majority of the distances is associated with improbable parameters under the posterior). All the above is independently repeated 30 times, so in the end for each bootstrap strategy we have a collection of $30\times 5$ percentiles of the ABC distances.
When using the stationary bootstrap, in order to make the data ``more stationary'', we apply order-two differences to both the observed and the simulated data. Then, corresponding summary statistics are obtained on such transformed data.

In Table \ref{tab:lv-boot-comparison} we report the medians of the percentiles across the thirty independent attempts. Recall for this case study each time series has length $n_{\mathrm{obs}}=32$. In Table \ref{tab:lv-boot-comparison}, the strategy denoted NOBB-fixed uses four non-overlapping blocks (hence each block has size 8), and the indeces that are sampled to form the 256 datasets from each prior-predictive simulation do not vary across the 1,000 prior predictive simulations. NOBB-rand instead assumes that such indeces do vary randomly across the 1,000 prior predictive simulations. OBB considers the case where four overlapping blocks of fixed size 8 are bootstrapped and finally the stationary bootstrap (SB) samples blocks of randomly varying size (the size is sampled from a uniform distribution over the interval [4,8]). We do not include the SB results in the table, not to trick the reader, and the reason is that, unfortunately, these cannot be readily compared with the other methods, since the distances in SB are computed on transformed data (order-two differences). So while the SB distances result smaller than the other methods this does not immediately translate in a better performance. We write the SB results here for the interested reader: 5.4 (0.1th perct.),  6.6 (0.2th perct.),  7.5 (0.3th perct.), 8.2 (0.4th perct.),  8.9 (0.5th perct.).

Among the methods in Table \ref{tab:lv-boot-comparison}, and at least for the considered dataset, the overlapping block bootstrap performs best since the ABC distances are on average smaller. It has to be added that the size of the blocks is known to have an impact on the performance, but for OBB we only considered blocks of size eight.
\begin{table}[]
    \centering
    \begin{tabular}{lccccc}
    \hline
        bootstrap method & 0.1th perct. & 0.2th perct. & 0.3th perct. & 0.4th perct. & 0.5th perct.\\
    \hline
       NOBB-fixed & 10.9 &  13.6 & 15.0 & 17.6  & 19.2\\
       NOBB-rand & 9.7 &  12.6 &  14.9 & 16.2  & 17.4\\
       OBB & 7.5 &   8.5 &  9.1  &  9.5  & 10.2\\
       \hline
    \end{tabular}
    \caption{Lotka-Volterra: medians of some (small) percentiles of prior-predictive ABC distances across 30 independent attempts, using several bootstrap methods. Smaller values denote better performance.}
    \label{tab:lv-boot-comparison}
\end{table}

\subsection{Setting strata for Lotka-Volterra}\label{sec:setting-lv-strata}

We can use percentiles  of the OBB distances, see Table \ref{tab:lv-boot-comparison}, to guide the construction of the strata. Of course, distances simulated from the prior predictive can get very large values, as opposed to distances corresponding to values of the parameters that are similar to those that generated the data (and for Lotka-Volterra, these can lie in a narrow region of the parameters space, as remarked in \citealp{papamakarios2016fast}). As we previously mentioned, we can assume that we have already reached high posterior probability regions, for example using some preliminary run via rABC-MCMC  (as for the g-and-k study) or other samplers. Since during ABC-MCMC we do not propose parameters from the prior, we set strata using percentiles that are even smaller than those in Table \ref{tab:lv-boot-comparison}:  $\mathcal{D}_1=(0,2.4)$, $\mathcal{D}_2=(2.4,4.8)$, $\mathcal{D}_2=(4.8,\infty)$ since the percentiles in Table \ref{tab:lv-boot-comparison} pertain simulations from the prior predictive distances, but with ABC-SMC we should manage to afford smaller percentiles.

\subsection{A computationally expensive Lotka-Volterra experiment}\label{sec:lv-expensive}

As mentioned in the main text, exact simulation of the Markov jump process is possible via the Gillespie algorithm. This algorithm produces a realization of the process at random times $t$, until a user defined maximal time $T_\mathrm{max}$ is reached. In our case this was set to $T_\mathrm{max}=64$, and the $n_{\mathrm{obs}}=32$ observations for each species resulted by linear interpolation of the produced process at a course grid of integer times [0,2,4,...,64]. In this section we want to consider a computationally more intensive simulation scenario, while still having 32 observations per species. Our intention is to produce a case study where the cost of data resampling and summary statistics computation is low compared to data simulation, which we believe is a more common scenario for realistic computer expensive models. Actual observations are the same ones as in the previous section, however for the inference part what we do here is prefixing the size $D$ of the time-mesh we use to simulate the Markov jump process generated by the Gillespie algorithm. We take  $D=620,001$ equispaced time points between time 0 and time 64 (i.e. a simulation timestep of $10^{-4}$), and then at each $\theta$ we run the Gillespie algorithm on $D$, and interpolate the resulting simulation grid at the usual  integer times [0,2,4,...,64] to obtain 32 values. We run the inference schemes for a total of 20,000 iterations and otherwise use the same setup as in the previous section, see results in Table \ref{tab:lv-intensive}. We do not re-execute ABC-SMC in this experiment given the high computational demands.  However for rsABC-MCMC we see that (i) inference is better than for pmABC-MCMC, see e.g. the credible interval for $\theta_1$ (and compare with ABC-SMC in Table 2 in the main text) and (ii) our stratified Monte Carlo version is between 11-13 times more efficient than pmABC-MCMC, in terms of ratios of ESS/min, depending on whether we compare with the $M=1$ or $M=2$ case. 
\begin{table}
\centering
\resizebox{\textwidth}{!}{  
\begin{tabular}{lccccccccc}
\hline
{} & $\theta_1$ & $\theta_2$ & $\theta_3$ & accept. rate & IAT & ESS & runtime & ESS/min \\
{} & {} & {} & {} & (\%) & {} & {} & (min) & \\
\hline
true parameters & 1 & 0.008 & 0.6 & & & &\\
pmABC-MCMC ($\delta=0.6$, $M=1$) & 0.988 [0.823,    1.190] &    0.008 [0.007, 0.011] &  0.629 [0.507, 0.873] & $\approx0.7\%$ & 747.3 & 15.2 & 189.7 & 0.08\\
pmABC-MCMC ($\delta=0.6$, $M=2$) & 1.013 [0.862,  1.231]  &  0.008 [0.007,   0.009]  &  0.618 [0.524, 0.721] & $\approx1\%$ & 597.9 & 27.0 & 374.6 & 0.07 \\
rsABC-MCMC ($\delta=4.8$) & 0.964 [0.708, 1.275]  &  0.009 [0.007, 0.012]  &  0.674 [0.501, 0.987] & $\approx 5\%$ & 77.6 & 201.1 & 227 & \textbf{0.89}\\
\hline
\end{tabular}
}
\caption{\footnotesize{Computer intensive Lotka-Volterra: Mean and 95\% posterior intervals for $\theta$ using several algorithms. Notice rsABC-MCMC uses a much larger $\delta$ than pmABC-MCMC, see main text. Numbers pertaining all ABC-MCMC strategies (including runtime) are based on the last 15,000 samples. The IAT reported is the largest autocorrelation time across chains and ESS is the smallest effective sample size across chains. The best performance is in bold (ESS/min for rsABC-MCMC is 11-13 times larger than the ESS/min for pmABC-MCMC).}}
\label{tab:lv-intensive}
\end{table}

\section{Self-tuned threshold in ABC-MCMC}\label{sec:tuning-abc}

Tuning ABC-MCMC algorithms is unfortunately not straightforward. Here we attempt a tuning strategy that we implemented only for the g-and-k study, see section \ref{sec:g-and-k}. For ABC-SMC samplers (or ABC-rejection for particularly simple problems), once an implementation is in place, they are easier to tune. We leave the specification of an ABC-SMC using stratified Monte Carlo for future research.

The possibility to simulate many artificial datasets at each proposed $\theta$ allows to tune the ABC threshold $\delta$, as the number of rABC-MCMC iterations increases. That is, we start rABC-MCMC at an initial parameter $\theta^0$, from which we simulate a first artificial dataset $x^*$. From $x^*$ we obtain $R$ resampled datasets $x^*_1,...,x^*_{R}$, and the corresponding statistics are $s^*_1,...,s^*_{R}$. We can then set an initial scaling matrix $\Sigma$ for these statistics, for example set $\Sigma=I_{n_s}$, the $n_s\times n_s$ identity matrix, which will be updated after an appropriate burnin. At this point, it is possible to compute the vector  of all distances $d=(d_1,...,d_{R})$, with $d_r=\sqrt{(s^{*}_r-s)'\Sigma^{-1}(s^{*}_r-s)}$, and obtain an initial threshold $\delta_0$ to be used for a number of iterations. A standard way to obtain $\delta_0$ is given by considering $\psi$-percentiles, which goes back to at least \cite{beaumont2002approximate} and has later been considered in different algorithms, see for example \cite{lenormand2013adaptive} and \cite{picchini2016accelerating}. In the $g$-and-$k$ case study (section \ref{sec:g-and-k}) we compute $\delta_0$ as the $\psi$-percentile of all distances $d$, with $\psi = 5$ (i.e. the 5th percentile). 

Once $\delta_0$ is obtained as above, $K$ burnin iterations of rABC-MCMC are executed. At iteration $K+1$, matrix $\Sigma$ is updated to be again diagonal, but with non-zero entries given by the squared median absolute deviation (MAD) of all summaries $s^*$ simulated up to iteration $K+1$. To this end we collect all simulated summaries (including those corresponding to rejected proposals) into a $(K\cdot R)\times n_s$ matrix, then for each column we obtain the corresponding MAD value. Finally, $\Sigma$ contains the $n_s$ squared MADs on its main diagonal. We do not further update $\Sigma$ for the remaining iterations. Instead, we periodically check whether it is appropriate to reduce the value of the threshold from $\delta_t$ to $\delta_{t+1}$, so that in the end we have a sequence of $T+1$ decreasing thresholds $\delta_0>\delta_1>\cdots >\delta_T$. 
 
We periodically checked (say every 5\% of the total number of rABC-MCMC iterations) whether the following two conditions were simultaneously satisfied: (i) the current proposal $\theta^*$ has been accepted and (ii) the summaries produced by the accepted $\theta^*$ are such that $\sum_{r=1}^{R} \mathbb{I}_{d_r<\delta} \geq 0.05\cdot R$ at that specific  iteration of rABC-MCMC. That is, if the number of distances that is smaller than the currently used $\delta_t$ is at least $5\%$ of the number of resamples, we lower the value of the threshold (again, provided that the proposed $\theta^*$ has been accepted). We use this criterion to avoid reducing $\delta_t$ when its value is apparently already small enough. When condition (i) is not satisfied at the iteration when we are supposed to check whether $\delta$ can be reduced, we do not wait until the next 5\% of rABC-MCMC iterations is processed: instead we check (ii) as soon as a proposal is accepted. 
Given the above, we attempt at reducing the threshold from its current value $\delta_t$ to $\delta_{t+1}$ according to the criterion
$\delta_{t+1}:=\min(\delta_{t},d_\psi)$,
where $d_\psi$ is the $\psi$-percentile of $d$.

\section{g-and-k distribution}\label{sec:g-and-k}

The g-and-k distribution is a standard toy model for ABC studies (\citealp{allingham2009bayesian,fearnhead-prangle(2011),picchini2018likelihood}), in that its simulation is straightforward, but it does not have a closed-form probability density function (pdf). Therefore the likelihood is analytically intractable. However, it has been noted in \cite{rayner2002numerical} that one can still numerically compute the pdf, by 1) numerically inverting the quantile function to get the cumulative distribution function (cdf), and 2)  numerically differentiating the cdf, using finite differences, for instance. Therefore it is possible to attain ``nearly exact'' (up to numerical discretization error) Bayesian inference via MCMC targeting $\pi(\theta|x)$ instead of $\pi(\theta|s)$. This approach is implemented in the \texttt{gk} R package \citep{gk}.

The  $g$-and-$k$ distributions is a flexibly shaped distribution that is used to model non-standard data
through a small number of parameters. It is defined by its quantile
function $F^{-1}(z;\theta)$, where $F^{-1}(z;\theta):[0,1]\rightarrow \mathbb{R}$ is given by 
\begin{equation}
F^{-1}(z;A,B,c,g,k)= A+B\biggl[1+c\frac{1-\exp(-g\cdot r(z))}{1+\exp(-g\cdot r(z))}\biggr](1+r^2(z))^kr(z)
\label{eq:g-k-inverse}
\end{equation}
where $r(z)$ is the $z$th standard normal quantile, $A$ and $B$ are location and scale parameters and $g$ and $k$ are related to skewness and kurtosis. Parameters restrictions are $B>0$ and $k>-0.5$.
An evaluation of \eqref{eq:g-k-inverse} returns a draw ($z$th quantile) from the $g$-and-$k$ distribution or, in other words, a sample from the g-and-k distribution can be easily simulated by tossing a standard Gaussian draw $r^*$, i.e. $r^*\sim \mathcal{N}(0,1)$ and then a single scalar draw from g-and-k is given by plugging $r^*$ in place of $r(z)$ in \eqref{eq:g-k-inverse}. We assume $\theta=(A,B,g,k)$ as parameter of interest, by noting that it is customary to keep $c$ fixed to $c=0.8$, see \cite{drovandi2011likelihood,rayner2002numerical}. We use the summaries $s(x)=(s_{A,x},s_{B,x},s_{g,x},s_{k,x})$ suggested in \cite{drovandi2011likelihood}:
\begin{align*}
s_{A,x}&=P_{50,x} & s_{B,x}&=P_{75,x}-P_{25,x},\\ 
s_{g,x}&=(P_{75,x}+P_{25,x}-2s_{A,x})/s_{B,x} & s_{k,x}&=(P_{87.5,x}-P_{62.5,x}+P_{37.5,x}-P_{12.5,x})/s_{B,x}
\end{align*}
where $P_{q,x}$ is the $q$th empirical percentile of $x$. That is $s_{A,x}$ and $s_{B,x}$ are the median and the inter-quartile range of $x$ respectively.

We now describe our simulations. Only for the case of exact Bayesian inference and for pmABC-MCMC we start simulations at ground-truth parameter values. 
We use the simulation strategy outlined above to generate data $x$, containing $n_\mathrm{obs}=2,000$ independent samples from the g-and-k distribution using parameters $\theta=(A,B,g,k)=(3,1,2,0.5)$. We place uniform priors on the log-parameters: $A\sim U(-30,30)$, $B\sim U(0,30)$, $g\sim U(0,30)$, $k\sim U(0,30)$. We use the \texttt{gk} package to report nearly exact Bayesian inference, providing a useful comparison with ABC inference: the package employs the adaptive MCMC strategy in \cite{haario2001adaptive} to propose parameters, however it does not return the acceptance rate, hence the ``NA'' in Table \ref{tab:gk}. We run overall 20,000 iterations and the resulting posterior marginals (discarding the first 10,000 draws as burnin) are given as ``exact Bayes'' in Figure \ref{g-and-k:all-posteriors}. Recall this inference is conditional on data $x$ not on $s(x)$. 

\begin{table}
\centering
\resizebox{\textwidth}{!}{  
\begin{tabular}{lccccccccc}
\hline
{} & $M$ & $R$ & {}  & {} & {} & & IAT & ESS &  accept. rate (\%)\\
\hline
true parameters & {} & {}  & $A=3$ & $B=1$ & $g=2$ & $k=0.5$\\
\hline
exact Bayes (full data) & -- & -- & 3.028 [2.974,3.084] &   1.102 [0.999,1.206] &  2.075 [1.952,    2.205]  &  0.477 [0.423,0.533] & 350.5 & 29.5 & NA\\
pmABC-MCMC & 500 & -- & 3.024 [2.644,3.417]  &  1.106 [0.072,2.419]  &  1.975 [1.694,2.289] & 0.559 [0.200,0.902] & 50.1 & 221.5 & 20\\
rsABC-MCMC & -- & 500 & 3.013 [2.896,3.118]  &  1.005 [0.692,1.403]  & 1.986 [1.614,2.320]  &  0.592 [0.372,0.856] & 47.0 & 244.4 & 15\\
xrsABC-MCMC & -- & 500 & 3.020 [2.927,3.111] &   1.009 [0.736,    1.315] &    1.986 [1.705,2.283] &    0.575 [0.413,0.734] & 48.4 & 222.0 & 11\\
\hline
\end{tabular}
}
\caption{\footnotesize{g-and-k: Mean and 95\% posterior intervals for $\theta$ using several algorithms. Notice ``exact Bayes'' is based on the entire dataset, while other approaches use summary statistics. Numbers pertaining all strategies are based on the last 10,000 samples where the same $\delta$ is used. The IAT reported is the largest autocorrelation time across chains and ESS is the smallest effective sample size across chains.}}
\label{tab:gk}
\end{table}

\begin{figure}
\centering
\includegraphics[scale=0.4]{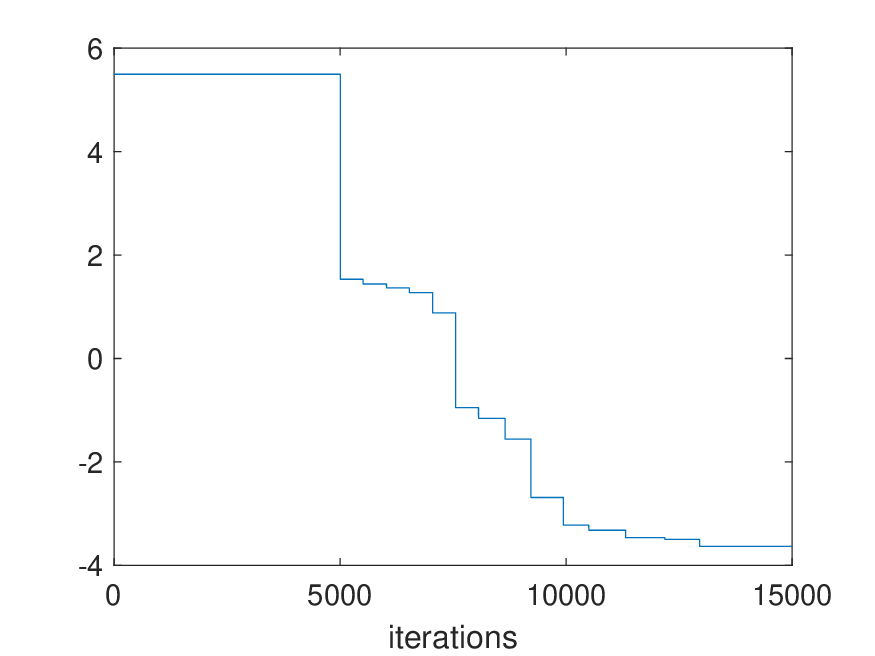}
\caption{g-and-k: evolution of the self-tuned $\log\delta_t$ during rABC-MCMC.}
\label{fig:g-and-k-logdelta}
\end{figure}

We now proceed at running rABC-MCMC. This is initialised at parameter values set far from the ground truth, namely starting parameter values (not log-transformed) are $\theta_0=(0.25,    2.72,  403.43,   10.00)$. We execute 15,000  iterations with rABC-MCMC using $R=500$ resampled datasets at each iteration. The parameters proposal kernel is the adaptive one from \cite{haario2001adaptive} with initial diagonal covariance matrix having variances $[0.1^2, 0.1^2, 0.1^2, 0.01^2]$, respectively for $\log A$, $\log B$, $\log g$ and $\log k$ (proposed parameters are generated on the log-scale). The proposal covariance is updated every 500 iterations. During rABC-MCMC the ABC threshold is automatically determined and updated as described in section \ref{sec:tuning-abc}, using $\psi =0.05$. The scaling matrix $\Sigma$ is obtained after the initial $K=5,000$ iterations, as described in section \ref{sec:tuning-abc}. The evolution of the threshold $\{\delta_t\}_{t=1}^T$ is in Figure \ref{fig:g-and-k-logdelta} (for ease of readability we report $\log \delta_t$). Its final value is $\delta_T=0.0264$, which is also used during the stratified rsABC-MCMC stage. The acceptance rate for rABC-MCMC at $\delta_T$ is around 12\%, which is way higher than typically desired with ABC algorithms when accurate inference is wanted, while values around 1--2\% have been shown to be often appropriate, e.g. \cite{fearnhead-prangle(2011)}.

\begin{figure}
\centering
\includegraphics[scale=0.7]{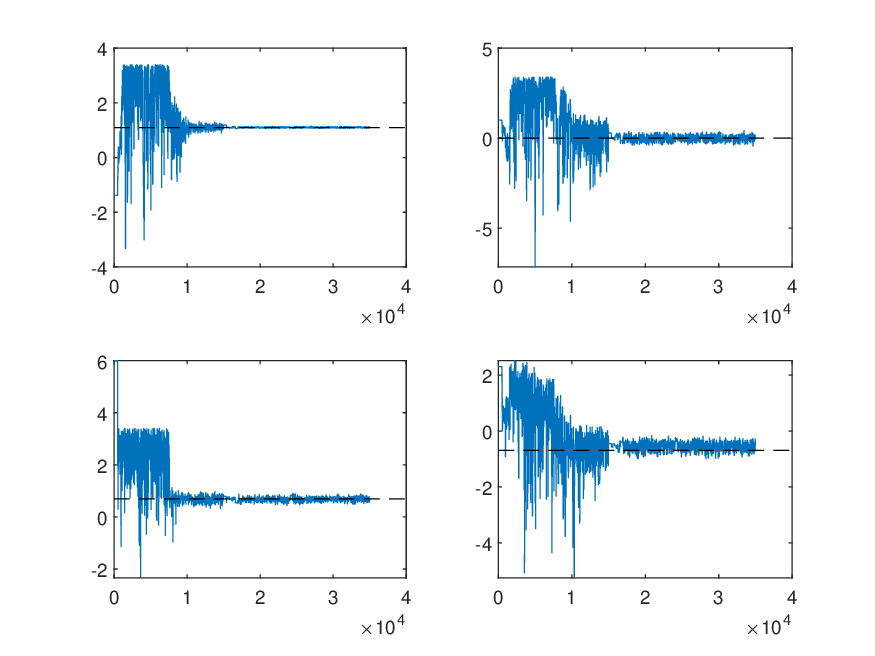}
\caption{g-and-k: chains for both rABC-MCMC (up to iteration 15,000) and xrsABC-MCMC (remaining 20,000 iterations). Upper-left $\log A$, upper-right $\log B$, bottom-left $\log g$, bottom-right $\log k$.  Dashed black lines are ground-truth parameter values.}\label{fig:g-and-k_abc_alliterations}
\end{figure}

\begin{figure}
\centering
\includegraphics[scale=0.6]{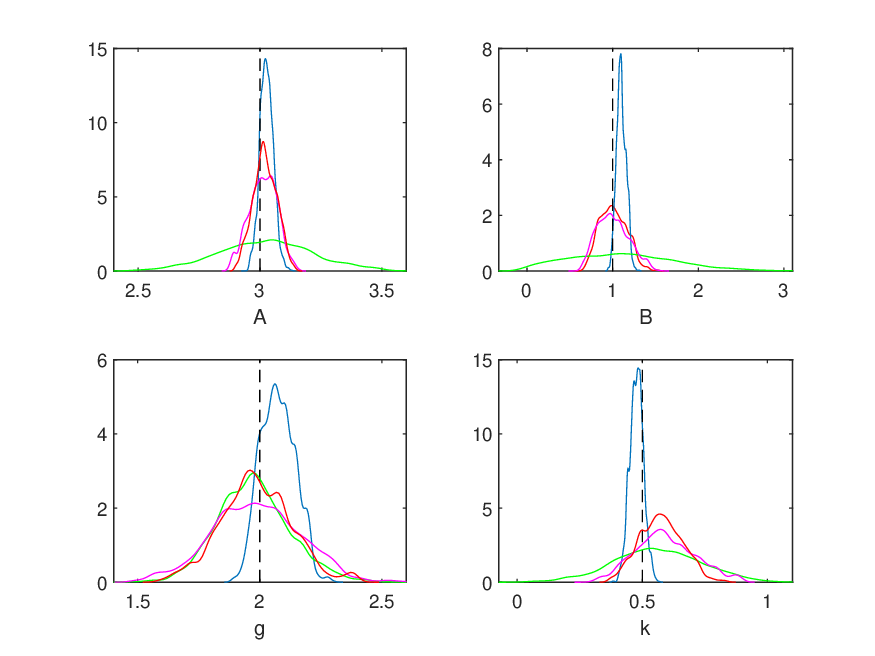}
\caption{\footnotesize{g-and-k: nearly exact posterior based on the full dataset (blue). Summaries-based pmABC-MCMC (green), rsABC-MCMC (magenta) and xrsABC-MCMC posteriors (red). All ABC marginals use the same $\delta=0.0264$. Dashed black lines denote ground-truth parameters.}}
\label{g-and-k:all-posteriors}
\end{figure}

Once rABC-MCMC has concluded, the xrsABC-MCMC procedure is initialized. We recall that xrsABC-MCMC is similar to rsABC-MCMC, except that $\bar{\mu}_{\mathrm{strat}}$ is employed in place of $\hat{\hat{\mu}}$, see section 4.2 in the main text. xrs-ABC-MCMC is initialized at the last accepted parameter value returned by rABC-MCMC, from which it inherits also the threshold $\delta=\delta_T=0.0264$ (which is kept constant throughout) and the covariance matrix used for the adaptive MCMC proposals generation. The stratified procedure comprises further 20,000 iterations using $R=500$. We consider three strata, namely $\mathcal{D}_1=\{s^* \text{ s.t. } ((s^{*}-s)'\Sigma^{-1}_*(s^{*}-s))^{1/2}\in (0,\delta_T/2]\}$,  $\mathcal{D}_2=\{s^* \text{ s.t. } ((s^{*}-s)'\Sigma^{-1}_*(s^{*}-s))^{1/2}\in (\delta_T/2,\delta_T]\}$, and $\mathcal{D}_3=\{s^* \text{ s.t. } ((s^{*}-s)'\Sigma^{-1}_*(s^{*}-s))^{1/2}\in (\delta_T,\infty)\}$. Here the acceptance rate is around 11\%. The chains for the entire simulation, including the rABC-MCMC stage, are in Figure \ref{fig:g-and-k_abc_alliterations}. Marginal posteriors for rABC-MCMC (using draws obtained with $\delta\equiv \delta_T$) and xrsABC-MCMC are in Figure \ref{g-and-k:all-posteriors}. 
Finally we run pmABC-MCMC with $M=500$. The simulation in this case is initiated at the ground truth parameter values, and we use the same scaling matrix $\Sigma$ as in rABC-MCMC and xrsABC-MCMC. Results for pmABC-MCMC in Figure \ref{g-and-k:all-posteriors} use the same threshold $\delta=0.0264$ as in previous analyses. The resulting acceptance rate is high, around 20\%, which is certainly not optimal for accurate inference, however the utility of the comparison in Figure \ref{g-and-k:all-posteriors} is that, for the same ABC threshold, we have that using resampling with stratification enhances the results considerably when considering a large $\delta$. For all parameters xrsABC-MCMC produces much more precise inference than pmABC-MCMC (it is not even necessary to produce plots for inference via rABC-MCMC, which is of course poorer than xrsABC-MCMC as evident from Figure \ref{fig:g-and-k_abc_alliterations}). Again, pmABC-MCMC could return better results if a much smaller $\delta$ was used, but that would require further iterations. Instead, we can instead keep a large value of $\delta$, have a large acceptance rate and still enjoy good results when using stratification.

\subsection{g-and-k: detail for the likelihood approximation}

\begin{figure}
\centering
\includegraphics[width=6cm,height=4cm]{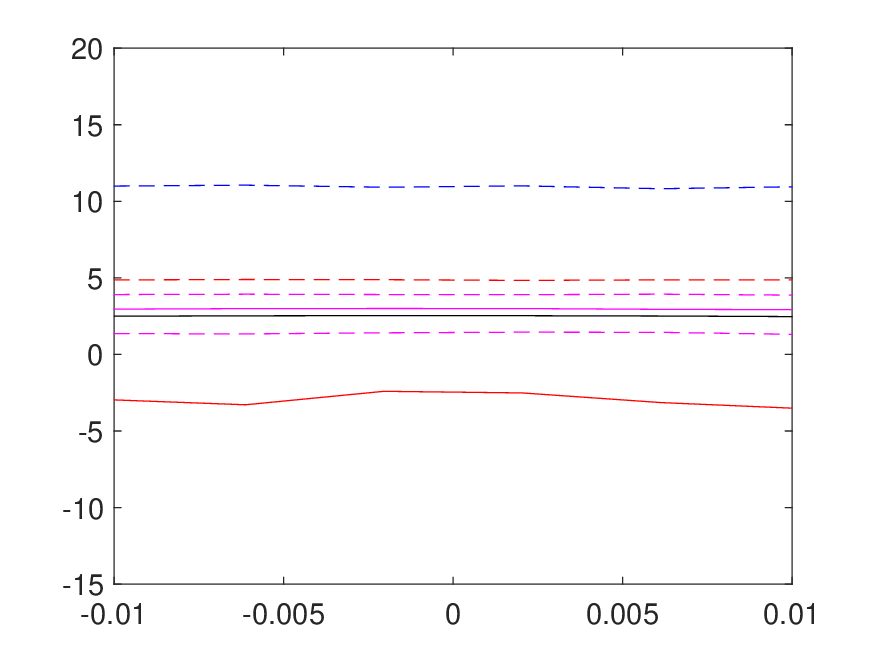}
\caption{\footnotesize{1D Gauss model: a magnified detail of Figure 2b in the main text. Exact loglikelihood of $s$ (black) and ABC loglikelihood estimated via standard ABC (red), rsABC (magenta), rABC (blue).}}\label{fig:gauss-three-likelihoods_zoomed}
\end{figure}

In Figure \ref{fig:gauss-three-likelihoods_zoomed} we show a zoomed-in version into the central part of Figure 2b as found in the main text. This reveals that both standard ABC and rABC returned a much more biased likelihood approximation, compared to rsABC. Notice the average of the rABC estimation is not even appearing in such plot since its bias is so large.

\end{document}